\documentclass[reprint,amsmath,amssymb,aps]{revtex4-2}
\usepackage{graphicx}
\usepackage{dcolumn}
\usepackage{bm}
\usepackage{xcolor}

\begin{document}
	
	\preprint{APS/123-QED}
	
	\title{Optical Appearances of Accreting Ellis-Bronnikov Wormholes Observed from Both Sides of Throats}
	
	\author{Yu-Hao Cui$^1$}
	\email{yuhaocui2025@163.com}
	
	\author{Kai Lin$^{2}$}
	\email{Contact author: lk314159@hotmail.com}
	
	\author{Sen Guo$^3$}
	\email{sguophys@126.com}
	
	\author{Yu Liang$^4$}
	\email{washy2718@outlook.com}
	
	\author{Tong Liu$^1$}
	\email{Contact author: tongliu@xmu.edu.cn}
	
	\affiliation{$^1$Department of Astronomy, Xiamen University, Fujian 361005, China\\
		$^2$Universidade Federal de Campina Grande, Campina Grande, PB 58429-900, Brasil\\
		$^3$College of Physics and Electronic Engineering, Chongqing Normal University, Chongqing 401331, China\\
		$^4$School of Big Data and Artificial Intelligence, Fuyang University of Technology, Anhui 236000, China}
	
	\date{\today}
	
	\begin{abstract}
		This study investigates the optical appearance of the Ellis-Bronnikov wormhole as viewed from both sides of its throat, under conditions of optically thick and thin accretion. By solving the geodesic equation, we derive the relationship between the impact parameter and the aiming distance of photons, and found that if the observer and the accretion disk are located on both sides of the throat, these two quantities are not equal. The optical image of the wormhole observed from the other side of the throat is obtained through the ray-tracing method. For optically thick accretion, increases in the parameter $n$ lead to an increase in the apparent size of the wormhole but a decrease in its brightness. For optically thin accretion, the image is similar to the internal and external inversion of the image observed from the other side. Furthermore, for optically thin accretion flows, the direct image does not block the emission from higher-order images, allowing radiation emitted from regions much closer to the event horizon to reach the observer. Our simulation results show that when the observer is on the $\mathcal{R}^+$ side, EB wormholes with small $n$ can mimic the images taken by the EHT to some extent, while wormholes with large $n$ or with the observer on the $\mathcal{R}^-$ side can be ruled out.
	\end{abstract}
	
	\maketitle
	
	\section{\label{sec:level1}INTRODUCTION}
	
	Compact objects in the universe can serve as probes for studying gravitational theories, among which black holes, as the most renowned compact objects predicted by general relativity, can already be imaged in accretion phases. In 2019, the Event Horizon Telescope (EHT) Collaboration released the first electromagnetic observation image of an accreting black hole, which was an important milestone in human astronomical observation and opened up a new way to understand compact celestial bodies such as black holes \cite{1,2,3,4,5,6}. Subsequently, the image of the supermassive black hole at the center of the Milky Way was also released \cite{7,8,9,10,11,12}. These remarkable achievements not only help us understand the spacetime characteristics around black holes and constrain modified gravity theories \cite{13}, but also provide information such as the magnetic field distribution around black holes \cite{14,15,16}. In recent years, there have been many studies on the shadows, photon rings and images of black holes under modified gravity\cite{17,18,19,20,21,22,23,23j1,23j2,23j3,23j4,23j5,23j6,23j7,23j8}.
	
	After the black hole was proposed, the concept of wormhole was also put forward \cite{24}. From the perspective of topology, wormholes can be regarded as tunnels that connect two regions, either within the same universe or between two distinct universes. In 1935, Einstein et al. attempted to eliminate the singularity of black holes by introducing a "bridge" at the event horizon that connects the external spacetime of the two Schwarzschild black holes \cite{25}. However, further studies on the Schwarzschild solution revealed that they failed to clearly distinguish between the coordinate singularity and the physical singularity \cite{26}. Among various types of wormholes, traversable wormholes have attracted the most scholarly attention due to their potential physical realizability and theoretical implications. In 1988, Morris and Thorne proposed the concept of traversable wormholes and proved that traversable wormholes necessarily violate the zero-energy condition \cite{27}. This condition can be resolved by introducing exotic matter or by modifying the theory of gravity \cite{28,29,30}. However, the latest research shows that there are also traversable wormholes in general relativity \cite{31}. 
	
	In this paper, we will focus on Ellis-Bronnikov (EB) wormhole, which is the first proposed traversable wormhole and independently discovered by Ellis \cite{32} and Bronnikov \cite{33} in 1973. In recent years, EB wormhole has been extended to modified gravity theories such as the Einstein-Æether theories \cite{34} and scalar tensor theories \cite{35}, and its stability has also been studied \cite{36,37}. The gravitational lensing effect of the EB wormhole has also aroused the interest of many scholars \cite{38,39}. An important work is the study by Bronnikov et al. on the deflection of light in asymmetric EB wormholes. They pointed out that if the photon sphere is located outside the throat, then there will be photons that pass through the throat and return \cite{40}.
	
	Although wormholes does not have the direct evidence of existence like black holes, its fascinating properties have attracted the attention of many scholars. Damour et al. discovered that many properties originally considered unique to black holes can also be simulated by wormholes, such as the no-hair theorem, event horizon and quasinormal modes \cite{41}. Cardoso et al. found only by conducting precise observations of the late ringdown signals can the subtle differences between the oscillations of wormholes and black holes be discovered \cite{42}. Tsukamoto et al. studied the differences between the Einstein rings of EB wormholes and black holes, and discussed the possibility of observing wormholes through astronomical observations \cite{43}. Therefore, the study of the differences between wormholes and black holes in optical images has great scientific value.

	After the release of the black hole image, work on the image of the wormhole has also been carried out gradually. One of the more significant advancements pertains to the wormholes constructed via the "cut and paste" method proposed by Visser \cite{44,45}. Wang et al.  conducted research on the asymmetric thin-shell wormhole (ATW) constructed by "cut and paste" method and found that when the spacetime on both sides of the throat is asymmetric, different shadows and photon rings from those of black holes will appear, which is caused by the reflection of photons by the double photon spheres \cite{46}. Based on this work, Peng et al.  studied the optical appearance of wormholes connected by two Schwarzschild spacetimes and found that there might be a gravitational lensing band between the two highly contracted photon rings in its image \cite{47}. Wielgus et al. studied the optical appearance of the ATW obtained by connecting two Reissner-Nordstrom spacetimes with different mass and charge parameters \cite{48}.
	
	Research on wormholes other than those in ATWs is also underway. Pual et al. used both semi-analytical methods and numerical ray tracing to study the optical appearance of Teo wormholes and Kerr-like wormholes under geometrically thin optical thick accretion. They found that the images of the wormhole accretion disks have significant differences compared to those of accreting black holes \cite{49}. Saleem et al. studied the temperature and radiation spectra of accretion disks in static spherically symmetric wormhole spacetimes, and found that the energy released by the accretion disks in the wormhole spacetime is greater than that in black hole spacetimes \cite{50}. Molla et al. studied the shadow of a new type of embedded wormholes in the context of general relativity, and found that the parameters of the wormhole have a significant impact on the size of the shadow. \cite{51}.
	
	Regarding the optical appearance of EB wormhole, two related papers were published two years ago. Huang et al. investigated the images of wormholes observed from both sides of the throat. However, their work did not take into account the case of optical thick accretion. \cite{53}. Ishkaeva et al. (2023) obtained the optical images of wormholes in the case of optical thick accretion, but they did not consider the situation of observing from the other side of the throat \cite{54}. Our work will adopt the original metric of the EB wormhole to study the optical appearance of the wormhole on both sides of the throat under the conditions of optically thick and optically thin accretion. The structure of this article is as follows: In Sec. \ref{sec:level2}, we introduce the Eliis-Bronnikov wormhole and analyze its global structure. In Sec. \ref{sec:level3}, We discussed the geodesics of wormhole spacetime. In Sec. \ref{sec:level4} and Sec. \ref{sec:level5}, we respectively study the optical appearance of wormholes in the two accretion backgrounds of optical thick and optical thin. A brief summary is made in Sec. \ref{sec:level6}.
	
	\section{\label{sec:level2}ELLIS-BRONNIKOV WORMHOLE SOLUTION}
	The well-known Ellis-Bronnikov wormhole metric can be expressed in the following form:
	\begin{equation}\label{eq1}
		ds^2=-f(r)dt^2+\frac{1}{f(r)}dr^2+\rho(r)(d\theta^2+\sin^2{\theta}d\varphi^2),
	\end{equation}
	where
	\begin{equation}
		f(r)=exp[-m\frac{\pi-2\arctan{\frac{r-m}{\sqrt{n^2-m^2}}}}{\sqrt{n^2-m^2}}],
	\end{equation}
	and
	\begin{equation}
		\rho(r)=\frac{r^2+n^2-2mr}{f(r)},
	\end{equation}
	in which $n$ and $m$ are two parameters of the wormhole solution, satisfying $n>m$. The range of coordinates are respectively $-\infty<t<+\infty$, $-\infty<r<+\infty$, $0<\theta\leq\frac{\pi}{2}$, and $0<\varphi<2\pi$.
	
	\begin{figure*}
		\centering
		\begin{minipage}{0.4\textwidth}
			\centering
			\includegraphics[width=1\textwidth]{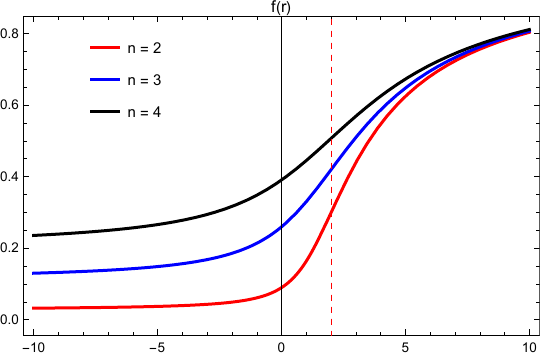}
		\end{minipage}
		\begin{minipage}{0.4\textwidth}
			\centering
			\includegraphics[width=1\textwidth]{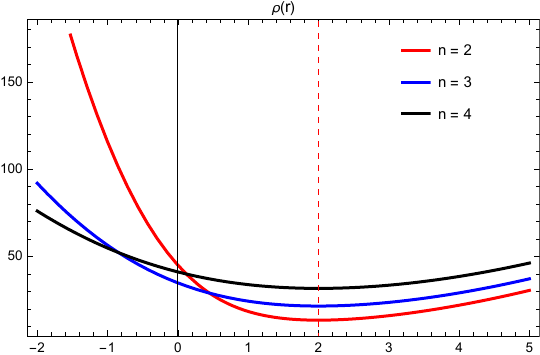}
		\end{minipage}
		\caption{Radial distributions of $f(r)$ and $\rho(r)$ with different parameters $n$. The red dashed line represents the radius of the throat.}
		\label{t1}
	\end{figure*}
	
	Upon calculation, one can found that the metric function $f(r)$ has the following asymptotic behavior \cite{54}:
	\begin{equation}
		\lim_{r\rightarrow+\infty}f(r)=1-\frac{2m}{r}+\mathcal{O}(r^{-2}),
	\end{equation}
	and
	\begin{equation}
		\lim_{r\rightarrow-\infty}f(r)=e^{\frac{-2\pi m}{\sqrt{n^2+m^2}}}(1-\frac{2m}{|r|})+\mathcal{O}(|r|^{-2}).
	\end{equation}
	
	It can be seen that the two spacetimes connected by the throat are both asymptotically flat. We will refer to the spacetime on the side where $r\rightarrow+\infty$ as $\mathcal{R}^+$, and the other side as $\mathcal{R}^-$. As $r\rightarrow \pm \infty $, according to the Minkowski metric, i.e.,
	\begin{equation}
		ds^2=-c^2dt^2+dr^2+r^2d\Omega^2,
	\end{equation}
	one can obtain that the speed of light in the two spacetimes is different, and there is:
	\begin{equation}
		\frac{c_{\mathcal{R}_+}}{c_{\mathcal{R}_-}}=exp(-\frac{\pi m}{\sqrt{n^2+m^2}}).
	\end{equation}
	
	Since the parameter $m$ can be analogized to the mass parameter in the Schwarzschild spacetime, in this paper we set $m=1$. Fig. \ref{t1} depicts the graph of $f(r)$ and $\rho(r)$ with different values of parameter $n$. From the figure, we can see that the throat of the wormhole is located at $r=2m$, which can be obtained by $\rho'(r)=0$. Additionlly, as $n$ increases, The difference between $\mathcal{R}^+$ and $\mathcal{R}^-$ will become increasingly smaller, and the two spacetimes will be increasingly flat.
	
	\section{\label{sec:level3}Geodesics}
	
	\subsection{Null geodesic}
	The $\theta$-geodesic equation in the spacetime described by metric (1) can be expressed as
	\begin{equation}
		\ddot{\theta}+2\frac{\rho'(r)}{\rho(r)}\dot{r}\dot{\theta}-\sin{\theta}\cos{\theta}\dot{\varphi}^2=0.
	\end{equation}
    
    It is straightforward to verify that $\theta=\pi/2, \dot{\theta}=0$ satisfies this equation identically. Therefore, we can confine their motion plane to the equatorial plane $(\theta=\pi/2,~ \dot{\theta}=0)$. The Lagrangian of photon motion is
	
	\begin{align}
		\mathcal{L} &=\frac{1}{2}g_{\mu \nu}\frac{dx^\mu}{d\lambda}\frac{x^\nu}{d\lambda}\nonumber \\
		&=\frac{1}{2}(-f(r)\dot{t}^2+\frac{\dot{r}^2}{f(r)}+\rho(r)\dot{\varphi}^2),
	\end{align}
	where $\lambda$ is affine parameter. It can be seen that $\frac{\partial \mathcal{L}}{\partial t}=0$ and $\frac{\partial \mathcal{L}}{\partial \varphi}=0$, so we can define the energy $E$ and angular momentum $L$ of photons as
	\begin{equation}
		E=\frac{\partial \mathcal{L}}{\partial \dot{t}}=-f(r)\dot{t}, \ \ L=\frac{\partial \mathcal{L}}{\partial \dot{\varphi}}=\rho(r)\dot{\varphi},
	\end{equation}
	in which the dots represent the derivatives of the affine parameter $\lambda$. According to the null geodesic equation $\mathcal{L}=0$, the radial motion equation can be expressed as:
	\begin{equation}\label{veff1}
		\dot{r}^2=E^2-f(r)\frac{L^2}{\rho(r)}.
	\end{equation}
	
	\begin{figure}
		\begin{minipage}{0.4\textwidth}
			\centering
			\includegraphics[width=1\textwidth]{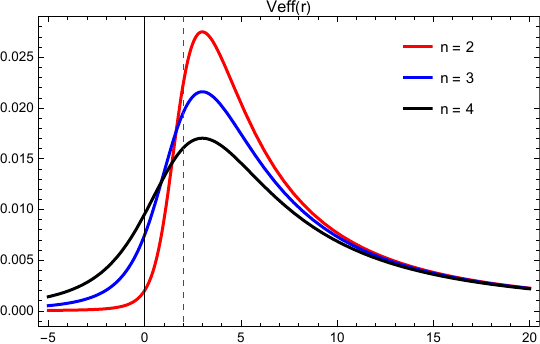}
		\end{minipage}
		\caption{Radial distributions of $V_{eff}(r)$ with different values of $n$. The red dashed line represents the radius of the throat.}
		\label{t2}
	\end{figure}
	
	By analogy with the definition of the potential function in Newtonian mechanics, we can define the effective potential $V_{eff}=f(r)\frac{L^2}{\rho(r)}$ for the radial motion of photons. Fig. \ref{t2} shows the distribution of the effective potential of the EB wormhole with different $n$. We can see from the figure that the effective potential has a unique peak. As $n$ increases, the peak of the effective potential gradually decreases. For different $n$, the radius corresponding to the peak is consistent. According to the condition:
	\begin{equation}
		V_{eff}'(r)=0,
	\end{equation}
	one can calculate that the peak is located at $r=3m$. That is to say, the radius of the circular orbit of the photon $r_c=3m$.

	In the next section, we mainly consider the observer at infinity. Here, we focus on the unbound orbits of photons. According to the behavior of the effective potential, we can classify the photons on unbound orbits as follows:
	
	For photons coming from the infinty of $\mathcal{R}^+$:
	\begin{itemize}
		\item $E>E_{max}$, the photon travels through the throat and eventually reaches the infinite  of $\mathcal{R}^-$;
		\item $E=E_{max}$, the photon gradually approaches the wormhole and eventually moves in a circular motion at $r=r_p$;
		\item $E<E_{max}$, the photon returns to infinity of $\mathcal{R}^+$ after passing through the turning point.
	\end{itemize}
	For photons coming from the infinty of $\mathcal{R}^-$:
	\begin{itemize}
		\item $E>E_{max}$, the photon travels through the throat and eventually reaches the infinite of $\mathcal{R}^+$;
		\item $E=E_{max}$, the photon travels through the throat and eventually moves in a circular motion at $r=r_p$;
		\item $E_{th}<E<E_{max}$, the photon returns to infinity of $\mathcal{R}^-$ after passing through the turning point;
		\item $E<E_{th}$, The photon does not reach the throat but returns to infinity of $\mathcal{R}^-$ after passing through the turning point.
	\end{itemize}
	where the turning point $r_p$ is the radius at which the photon's radial velocity becomes zero, $E_{max}$ represents the maximum value of the effective potential attained at $r=r_c$, and $E_{th}$ and $r_{th}$ correspond to the effective potential at the throat and the throat radius, respectively.
	
\begin{figure}
		\begin{minipage}{0.45\textwidth}
			\centering
			\includegraphics[width=1\textwidth]{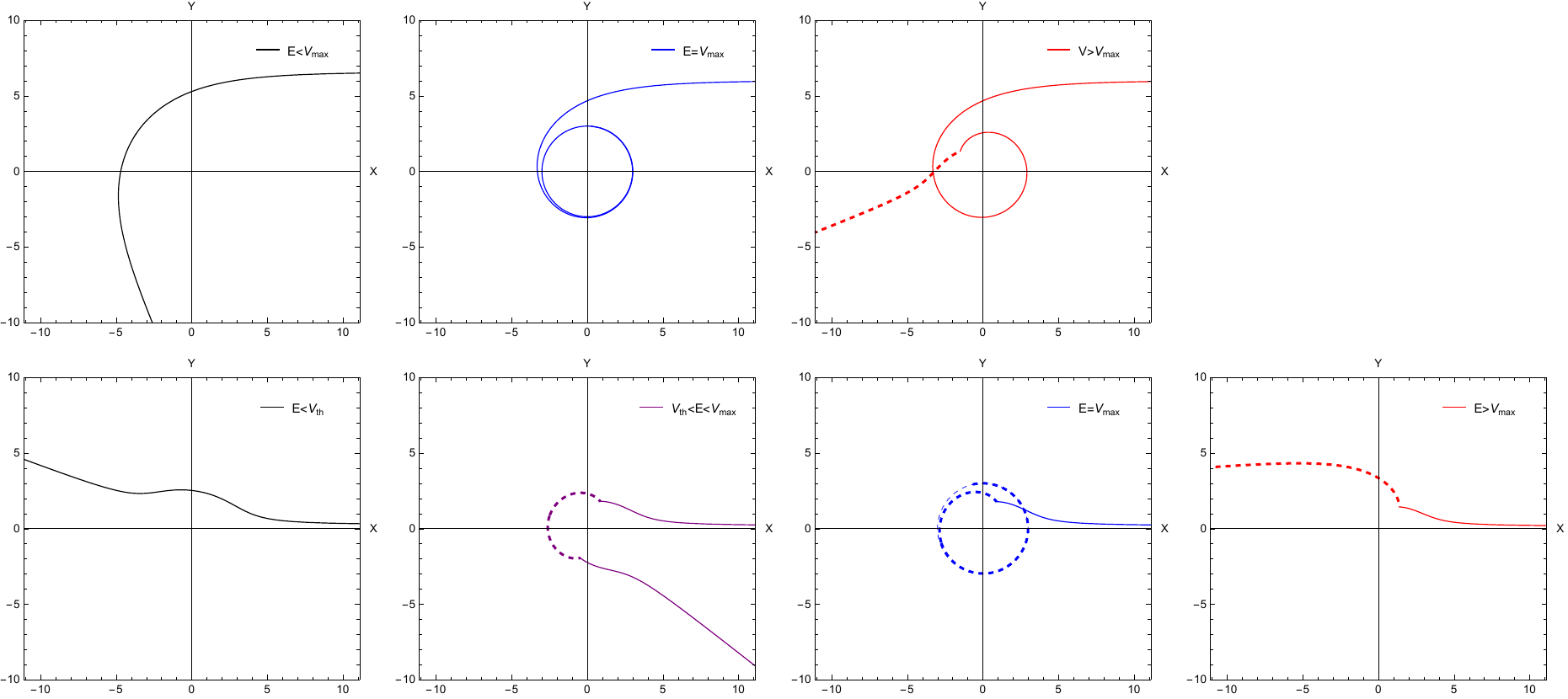}
		\end{minipage}
		\caption{Seven kinds of unbound trajectories of photons coming from the infinity of $\mathcal{R}^+$ and $\mathcal{R}^-$. Top panel: Photons from the infinity of $\mathcal{R}^+$. Bottom panel: Photons from the infinity of $\mathcal{R}^-$. The dashed lines represent the trajectory after passing through the throat.}
		\label{t3}
	\end{figure}

	To present the trajectories of these photons more intuitively, we eliminate $\lambda$ from Eq. (\ref{veff1}) and obtain the orbit equation:
	\begin{equation}
		\left(\frac{dr}{d\varphi}\right)^2=\rho(r)\left(\frac{\rho(r)}{b^2}-f(r)\right),
	\end{equation}
	where $b=\frac{L}{E}$ is called the impact parameter. Fig. \ref{t3} shows seven kinds of light trajectories originating from the infinity of $\mathcal{R}^+$ and $\mathcal{R}^-$ respectively. In order to draw the circular orbit of photons, we first calculate their critical impact parameter $b_c$, which can be obtained by solving the equations $V_{eff}=E^2$ and $V'_{eff}=0$:
	\begin{equation}
		b_c=exp[\frac{m(\pi-2arctan(\frac{2m}{\sqrt{n^2-m^2}}))}{\sqrt{n^2-m^2}}]\sqrt{n^2+3m^2}.
	\end{equation} 
	
\begin{figure*}
		\begin{minipage}{0.3\textwidth}
			\centering
			\includegraphics[width=1\textwidth]{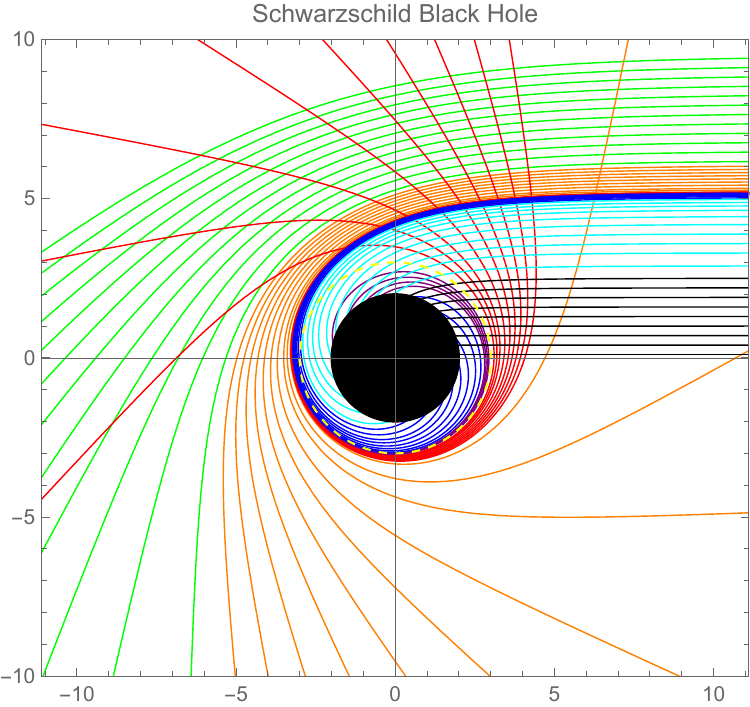}
		\end{minipage}
		\begin{minipage}{0.3\textwidth}
			\centering
			\includegraphics[width=1\textwidth]{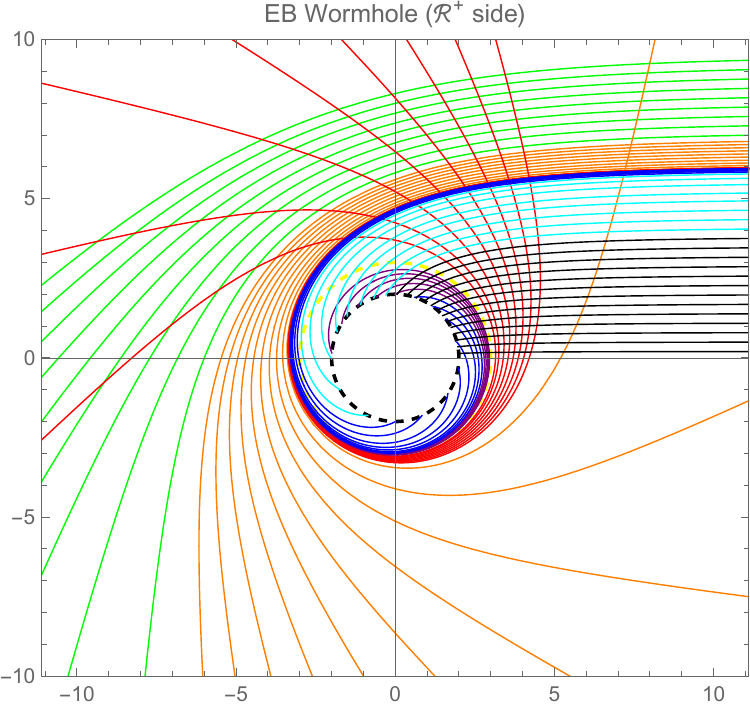}
		\end{minipage}
		\begin{minipage}{0.3\textwidth}
			\centering
			\includegraphics[width=1\textwidth]{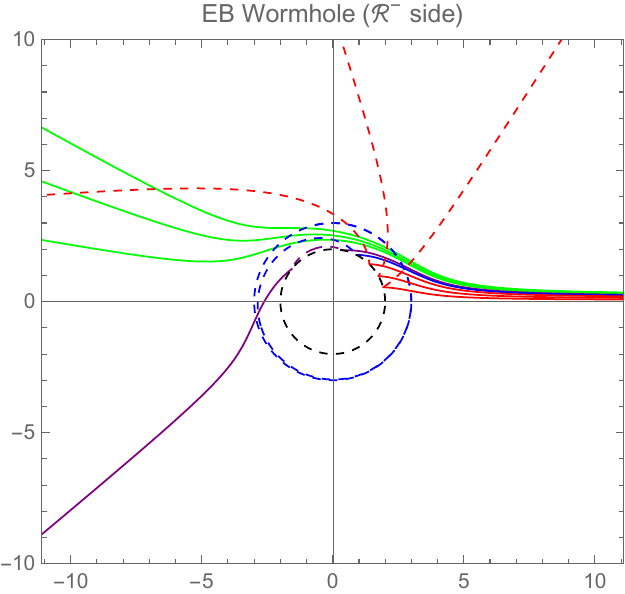}
		\end{minipage}
		\caption{The unbound trajectory of photons in EB wormhole($n=2$) spacetime. The black disk in the left picture represents the interior of the event horizon. The black dashed circular ring in the middle picture and right picture represents the wormhole's throat. Left panel: Schwarzschild black hole. Middle panel: EB wormhole(Photons from the infinity of $\mathcal{R}^+$). Right panel: EB wormhole(Photons from the infinity of $\mathcal{R}^-$).}
		\label{t4}
	\end{figure*}

	Fig. \ref{t4} illustrates the light ray trajectories around the Schwarzschild black hole and the EB wormhole. In the left andmiddle panels, the light rays are categorized according to the classification scheme in \cite{19}. The green and cyan curves correspond to trajectories with deflection angles in the interval $(0, 3\pi/2)$, the orange and blue curves represent those with deflection angles in $(3\pi/2, 5\pi/2)$, and the red and purple curves denote trajectories with deflection angles greater than $5\pi/2$. In the process of drawing the right panel, we note that for photons arriving simultaneously from infinity on both sides of the throat, the ratio of their aiming distances in the $\mathcal{R}^+$ and $\mathcal{R}^-$ spacetimes is fixed. Therefore, in the Appendix. \ref{app}, we discuss the relationship between the impact parameter $b$ and the aiming distance $d^*$, and find that for EB wormhole, there is:
	\begin{equation}\label{bwuli}
		b=\lim_{r\rightarrow \pm \infty}\frac{d^*}{f(r)}.
	\end{equation}
	
	The detailed derivation process of Eq. (\ref{bwuli}) can be found in the Appendix. \ref{app}. Suppose the observer is located at infinity in $\mathcal{R}^-$, then $b=exp(\frac{2\pi}{\sqrt{3}}) d^* \approx 37.6224 d^*$ for $n=2$. The right panel in Fig. \ref{t4} shows the trajectory of the light rays around the EB wormhole for photons coming from the infinity of $\mathcal{R}^-$. To avoid overcrowding caused by Eq. (\ref{bwuli}), only a selected set of rays is displayed. Among these, the green, purple, blue, and red lines correspond to photon trajectories with energies $E<E_{max}$, $E_{th}<E<E_{max}$, $E=E_{max}$, and $E>E_{max}$ respectively. 
	
	\subsection{Time-like geodesic}
	In order to study the motion of particles on the accretion disk, we investigate the timelike geodesics in this subsection. For a massive particle, the Lagrangian for free motion is
	\begin{align}
		\mathcal{L}_m &=\frac{1}{2}g_{\mu \nu}\frac{dx^\mu}{d\tau}\frac{x^\nu}{d\tau} \nonumber \\
		&=\frac{1}{2}(-f(r)\dot{t}^2+\frac{\dot{r}^2}{f(r)}+\rho(r)\dot{\varphi}^2),
	\end{align}
	where $\tau$ is proper time. According to the Noether's theorem, the energy $E_m$ and angular momentum $L_m$ of a massive particle can be defined as
	\begin{equation}
		E_m=\frac{\partial \mathcal{L}}{\partial \dot{t}}=-f(r)\dot{t}, \ \ L_m=\frac{\partial \mathcal{L}}{\partial \dot{\varphi}}=\rho(r)\dot{\varphi},
	\end{equation}
	where dots represent the derivatives of the proper time. According to the geodesic equation $\mathcal{L}=-1/2$, the radial motion equation of a massive particle can be expressed as
	\begin{equation}\label{lizi}
		\dot{r}^2=E_m^2-f(r)\left(\frac{L_m^2}{\rho(r)}+1\right).
	\end{equation}
	
	Taking the derivative of Eq. (\ref{lizi}) with respect to $\tau$, we can obtain the expression of radial acceleration:
	\begin{equation}\label{21}
		\ddot{r}=\frac{f(r)\rho'(r)L_m^2-f'(r)\rho(r)(\rho(r)+L_m^2)}{\rho(r)^2}.
	\end{equation}
	
	We begin by analyzing the scenario of $L_m = 0$, which corresponds to purely radial motion. In this case, the equation of motion simplifies to $\ddot{r} = -f'(r)$. The form of this acceleration indicates that particles are attracted toward the wormhole in the $\mathcal{R}^+$ spacetime but are repelled from it in the $\mathcal{R}^-$ spacetime. It follows that the accretion disk can only form on the $\mathcal{R}^+$ side. Given this, our subsequent analysis focuses solely on particle dynamics within $\mathcal{R}^+$. Particles on the accretion disk can be approximately regarded as particles in circular orbits, and the condition for a particle to move along a circular orbit is
	\begin{equation}\label{22}
		\dot{r}=0, ~ \ddot{r}=0.
	\end{equation}
	Substituting Eq. (\ref{22}) into Eqs. (\ref{lizi}) and (\ref{21}), the energy and angular momentum of the circular orbit example can be obtained as
	\begin{equation}
		\begin{split}
			E_m=\frac{f(r)\sqrt{\rho'(r)}}{\sqrt{f(r)\rho'(r)-f'(r)\rho(r)}}, \\ L_m=\frac{\rho(r)\sqrt{f'(r)}}{\sqrt{f(r)\rho'(r)-f'(r)\rho(r)}}.
		\end{split}
	\end{equation}
	
	Following the definition of the effective potential for photons, the effective potential $V_m(r)$ for particles can be defined as
	\begin{equation}
		V_m(r)=f(r)\left[\frac{L_m^2}{\rho(r)}+1\right].
	\end{equation}
	
	According to the condition for stable circular orbits $V_m''(r)\leq 0$, we can derive that the stable circular orbits exist in the range of $r \geq 4m+\sqrt{4m^2+n^2}$. Therefore, the innermost stable circular orbit in $\mathcal{R}^+$ spacetime is
	\begin{equation}
		r_{isco}=4m+\sqrt{4m^2+n^2}.
	\end{equation}
	
\begin{figure*}
		\begin{minipage}{0.4\textwidth}
			\centering
			\includegraphics[width=1\textwidth]{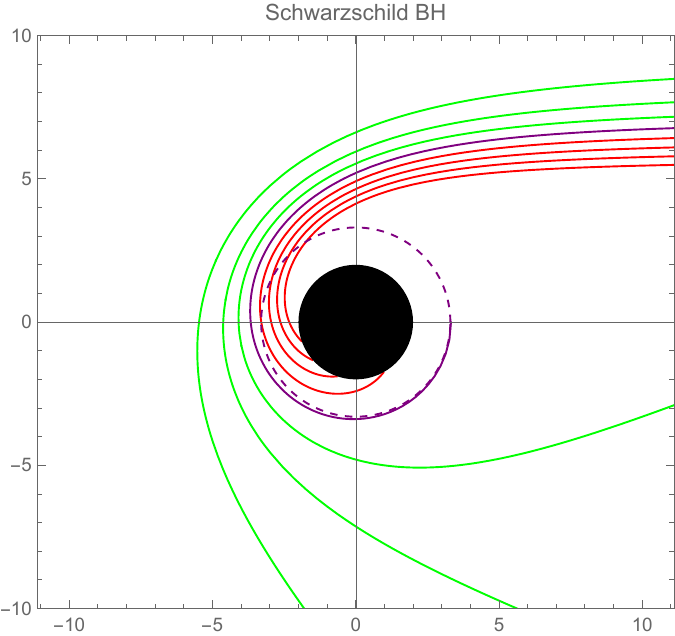}
		\end{minipage}
		\begin{minipage}{0.4\textwidth}
			\centering
			\includegraphics[width=1\textwidth]{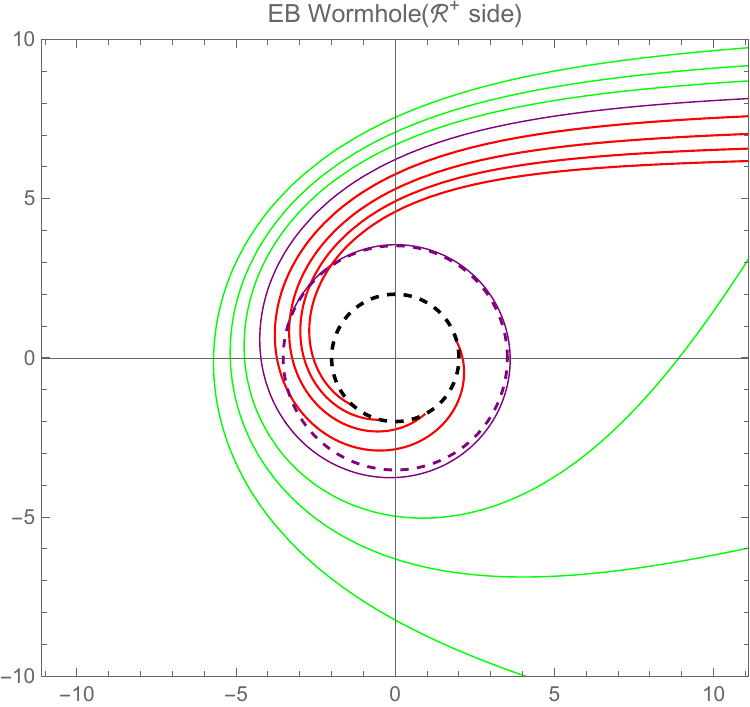}
		\end{minipage}
		\caption{The trajectory of particles around Schwarzschild and EB wormhole($n=2$). The black disk in the left panel represents the interior of the event horizon. The black dashed circular ring in the right panel represents the wormhole's throat. Left panel: Schwarzschild black hole. Right panel: EB wormhole(Photons from the infinity of $\mathcal{R}^+$). }
		\label{tjia}
	\end{figure*}

	In order to more clearly illustrate the motion of particles around an EB wormhole, we plot their trajectories in Fig. \ref{tjia}, in which the green lines indicate particles deflected by the black hole (wormhole); the purple lines represent particles in orbiting motion; and the red lines show particles falling into the horizon (throat).
	
	\section{\label{sec:level4}The optical appearance of the optically thick accretion disk}
	\subsection{Ray-tracing}
\begin{figure}
		\begin{minipage}{0.45\textwidth}
			\centering
			\includegraphics[width=1\textwidth]{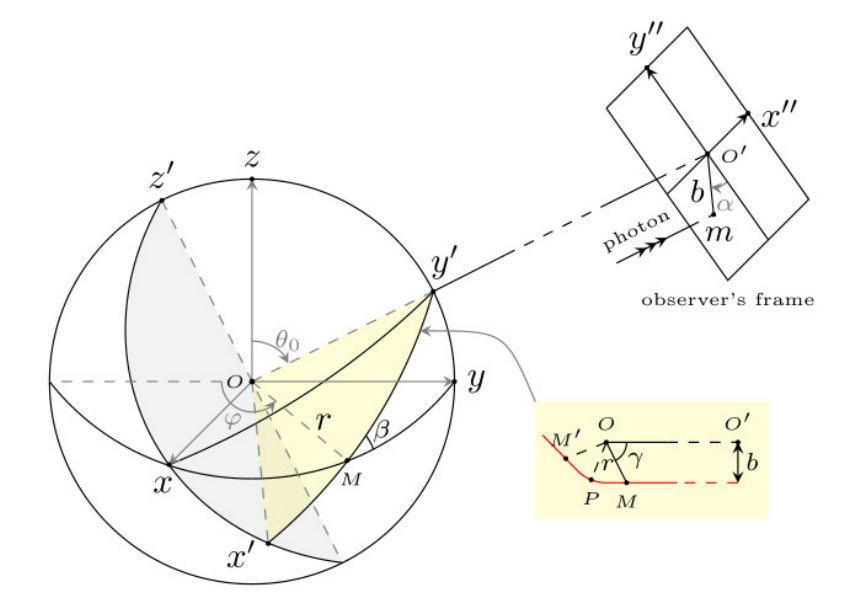}
		\end{minipage}
		\caption{The schematic diagram of ray-tracing \cite{56}.}
		\label{t5}
	\end{figure}

	In this subsection, we adopt the method proposed by Luminet \cite{18} to image the accretion disk of EB wormhole. The schematic diagram of ray-tracing is shown in Fig. \ref{t5} The origin is located at point $O$, and the accretion disk is situated on the $xoy$ plane. The photon starts from point $M$ on the accretion disk and eventually reaches point $m$ on the observation plane $x''o'y''$, with a deflection angle of gamma during this process. Based on the analysis in the previous section, the light emitted by the accretion disk will not return to the $\mathcal{R}^+$ spacetime after passing through the throat. Therefore, when the observer is in the $\mathcal{R}^+$ spacetime, the situation of the ray-tracing of EB wormhole is exactly the same as that of a black hole. For the situation of the observer in $\mathcal{R}^-$, we can observe that during the photon's movement, the deflection angle of the photon remains $\gamma$. Additionally, the geometric relationship shown in Fig. \ref{t5} still holds true. In conclusion, here we employ the ray-tracing method shown in Fig. \ref{t5} to study the optical appearance of EB wormholes with optically thick accretion.
	
	By applying the sine the sine theorem of spherical triangles, we can eliminate the auxiliary angle $\beta$ in Fig. \ref{t5} and establish the geometric relationship between the photon deflection angle $\gamma$ and the polar angle $\alpha$ of point $m$:
	\begin{equation}\label{24}
		\cos{\alpha}=\cos{\gamma}\sqrt{\cos^2{\alpha}+\cot^2{\theta_0}},
	\end{equation}
	where $\theta_0$ denotes the angle between the observation plane and the $z$-axis.
	
	\begin{figure*}
		\begin{minipage}{0.3\textwidth}
			\centering
			\includegraphics[width=1\textwidth]{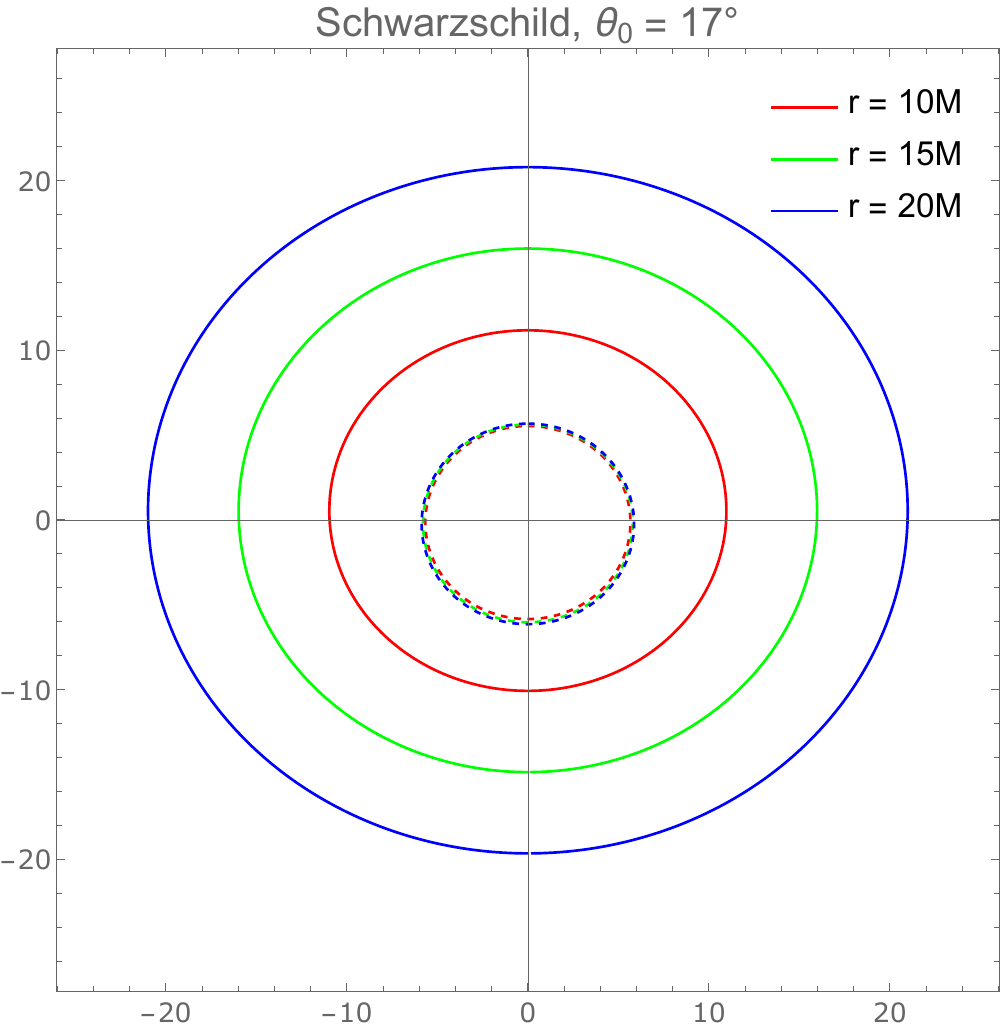}
		\end{minipage}
		\begin{minipage}{0.3\textwidth}
			\centering
			\includegraphics[width=1\textwidth]{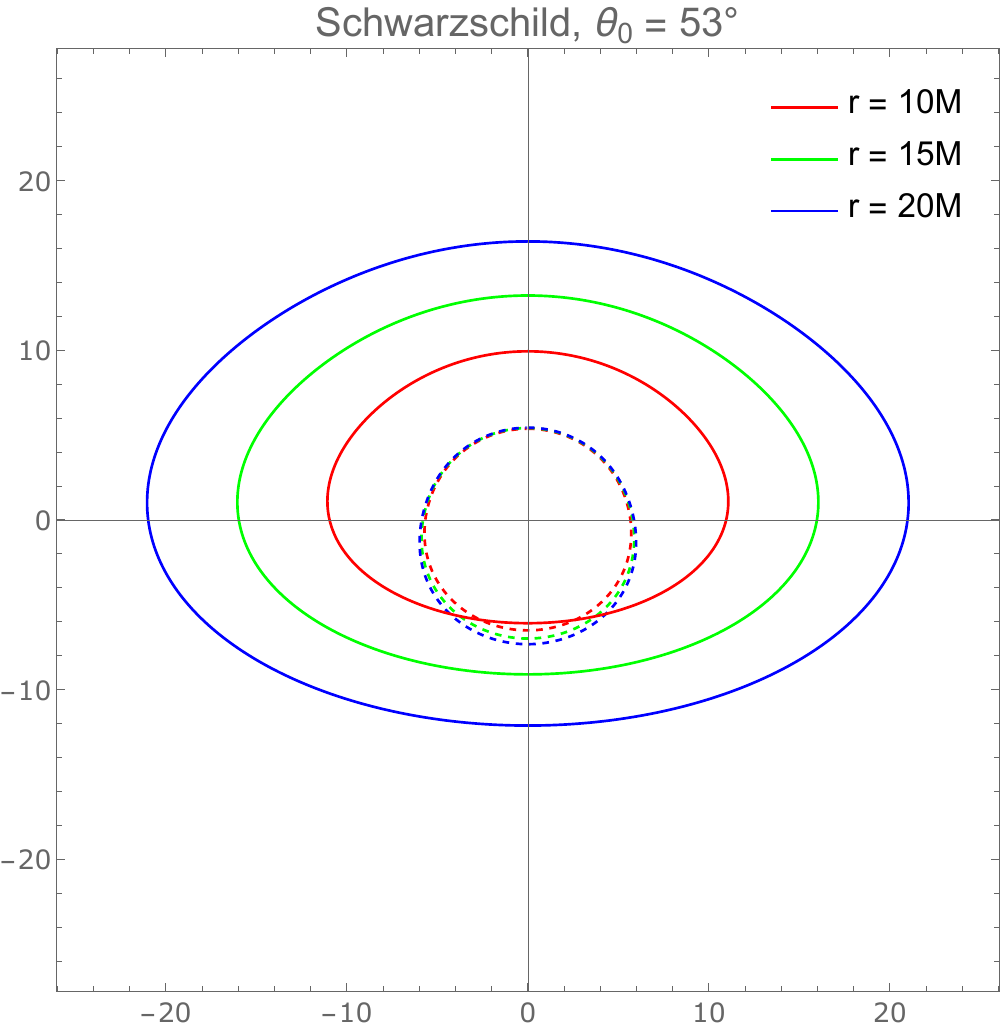}
		\end{minipage}
		\begin{minipage}{0.3\textwidth}
			\centering
			\includegraphics[width=1\textwidth]{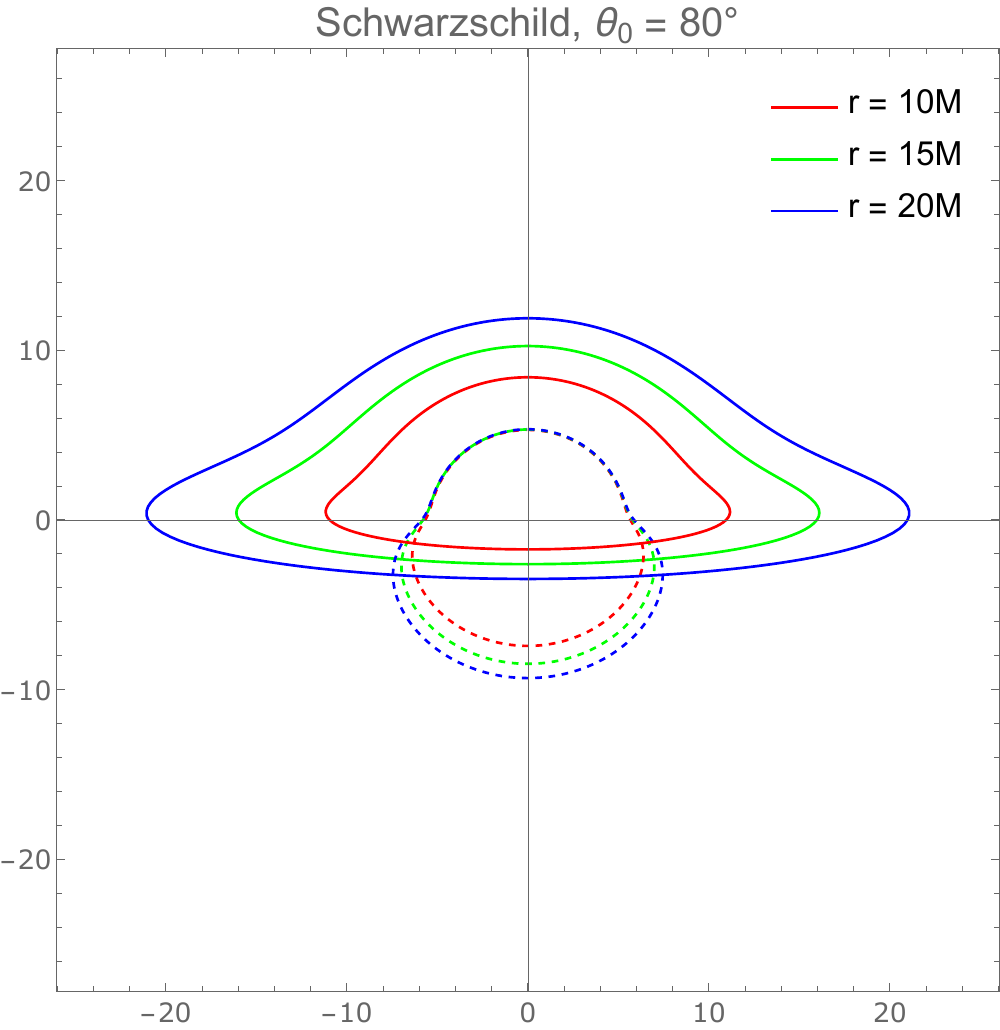}
		\end{minipage}
		\begin{minipage}{0.3\textwidth}
			\centering
			\includegraphics[width=1\textwidth]{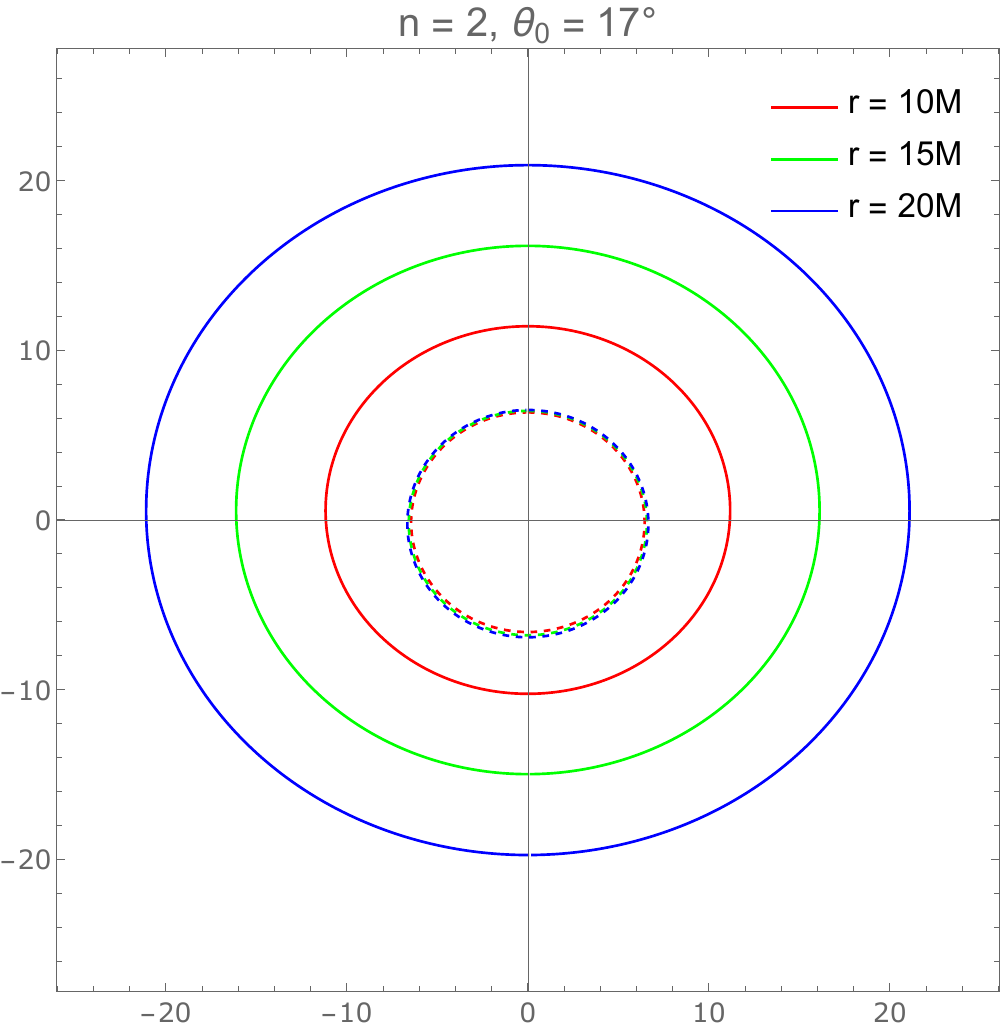}
		\end{minipage}
		\begin{minipage}{0.3\textwidth}
			\centering
			\includegraphics[width=1\textwidth]{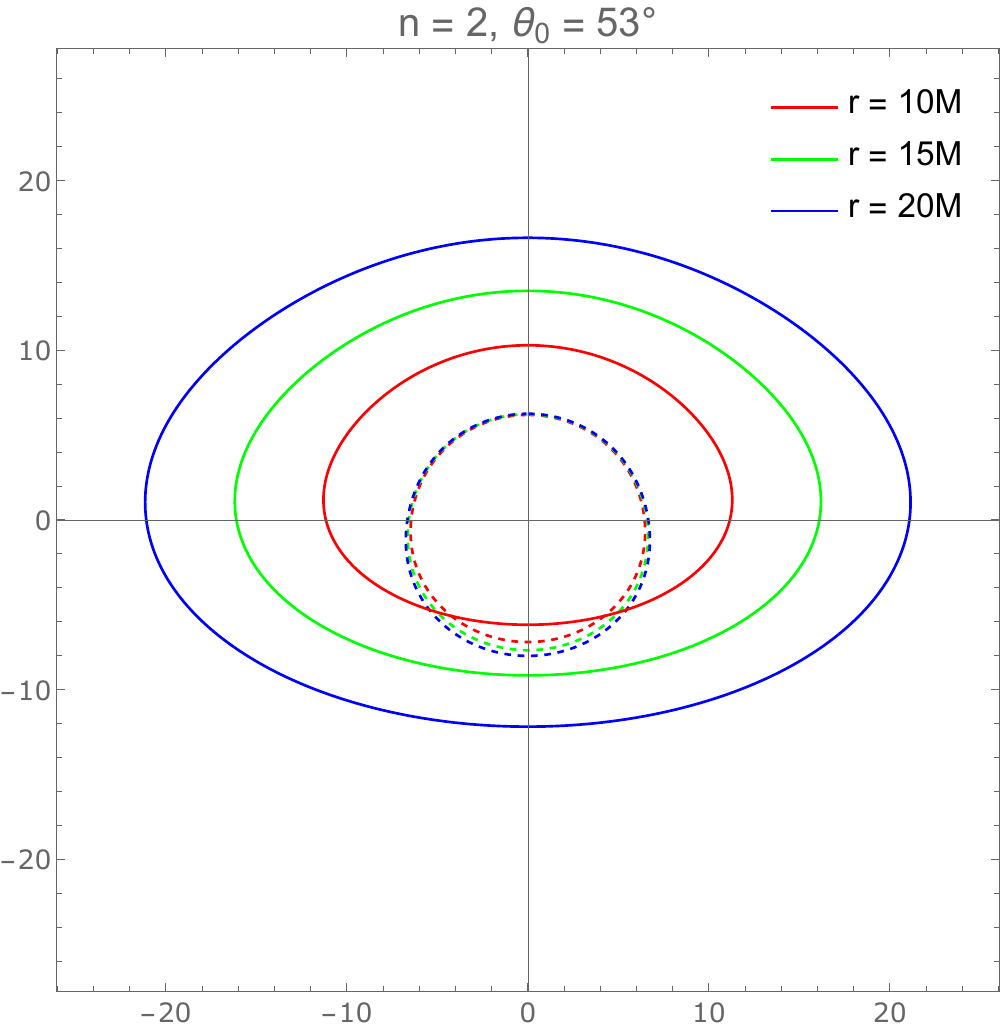}
		\end{minipage}
		\begin{minipage}{0.3\textwidth}
			\centering
			\includegraphics[width=1\textwidth]{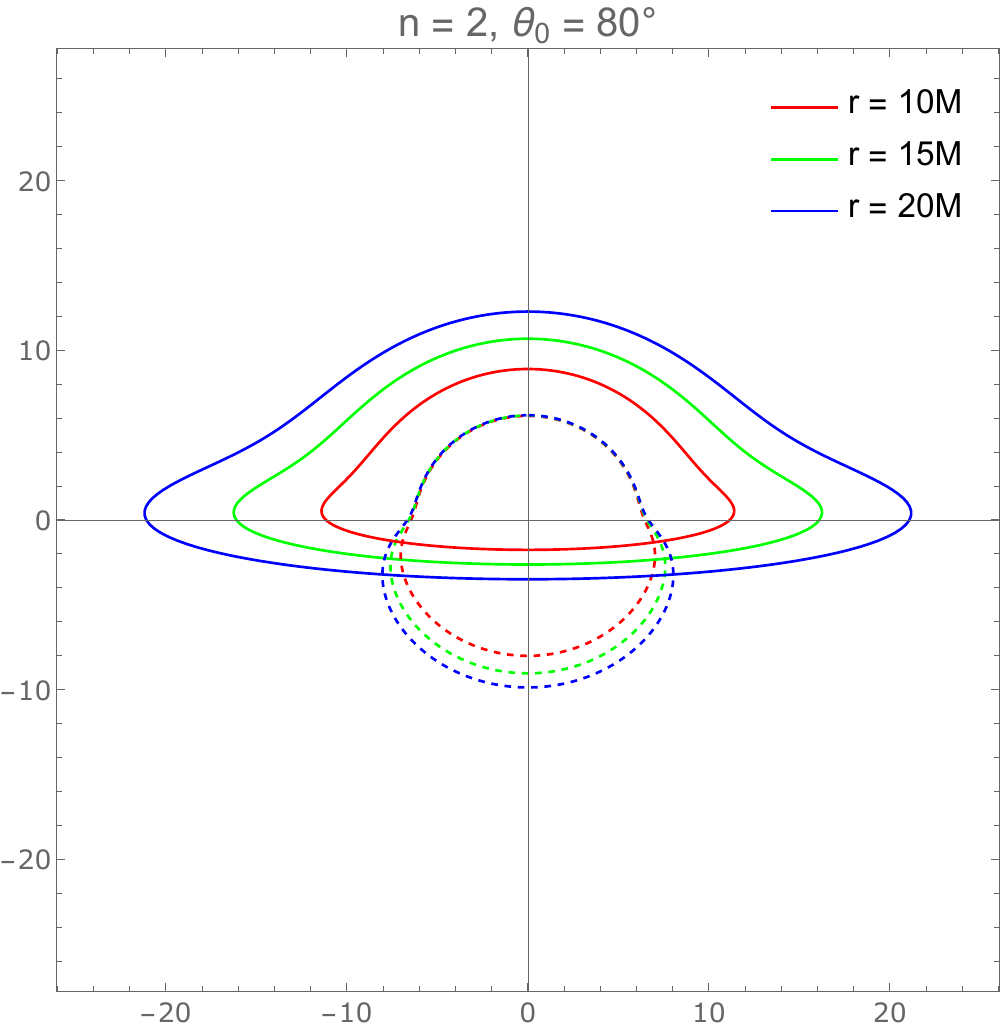}
		\end{minipage}
		\begin{minipage}{0.3\textwidth}
			\centering
			\includegraphics[width=1\textwidth]{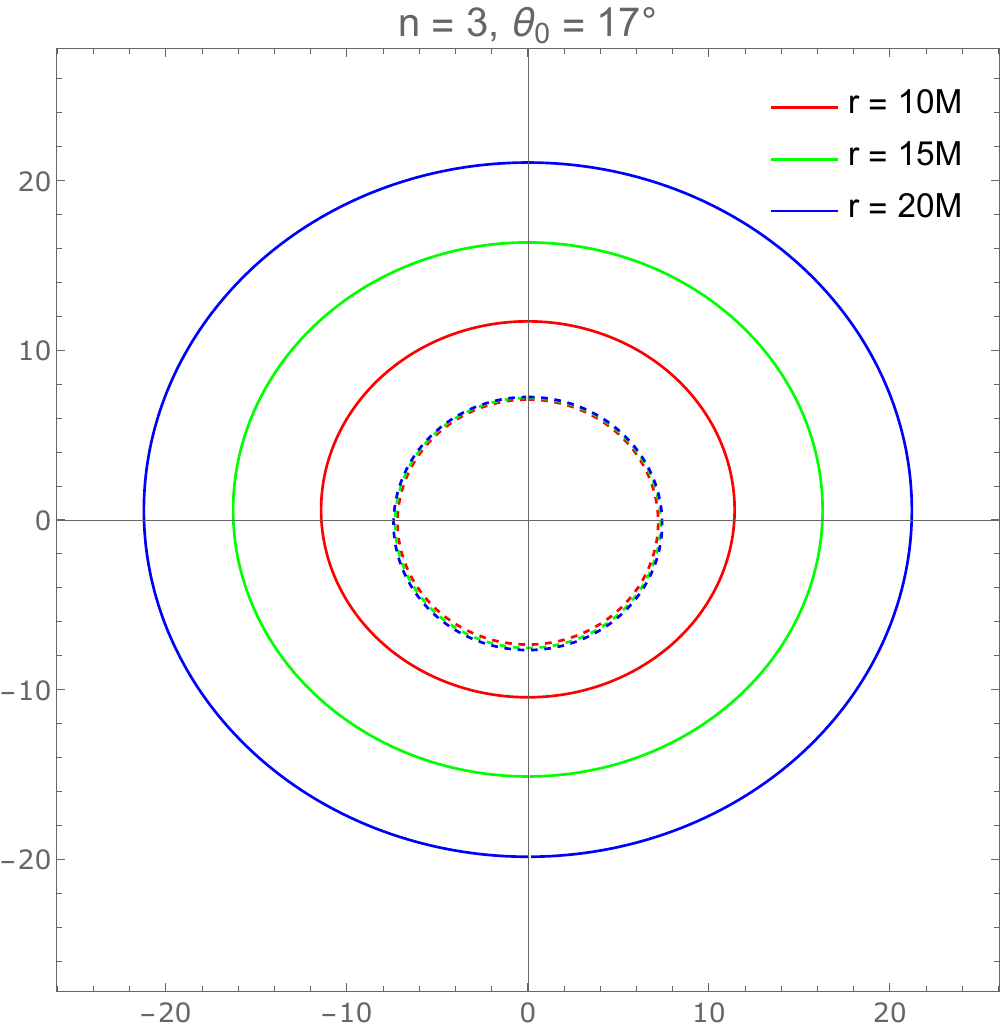}
		\end{minipage}
		\begin{minipage}{0.3\textwidth}
			\centering
			\includegraphics[width=1\textwidth]{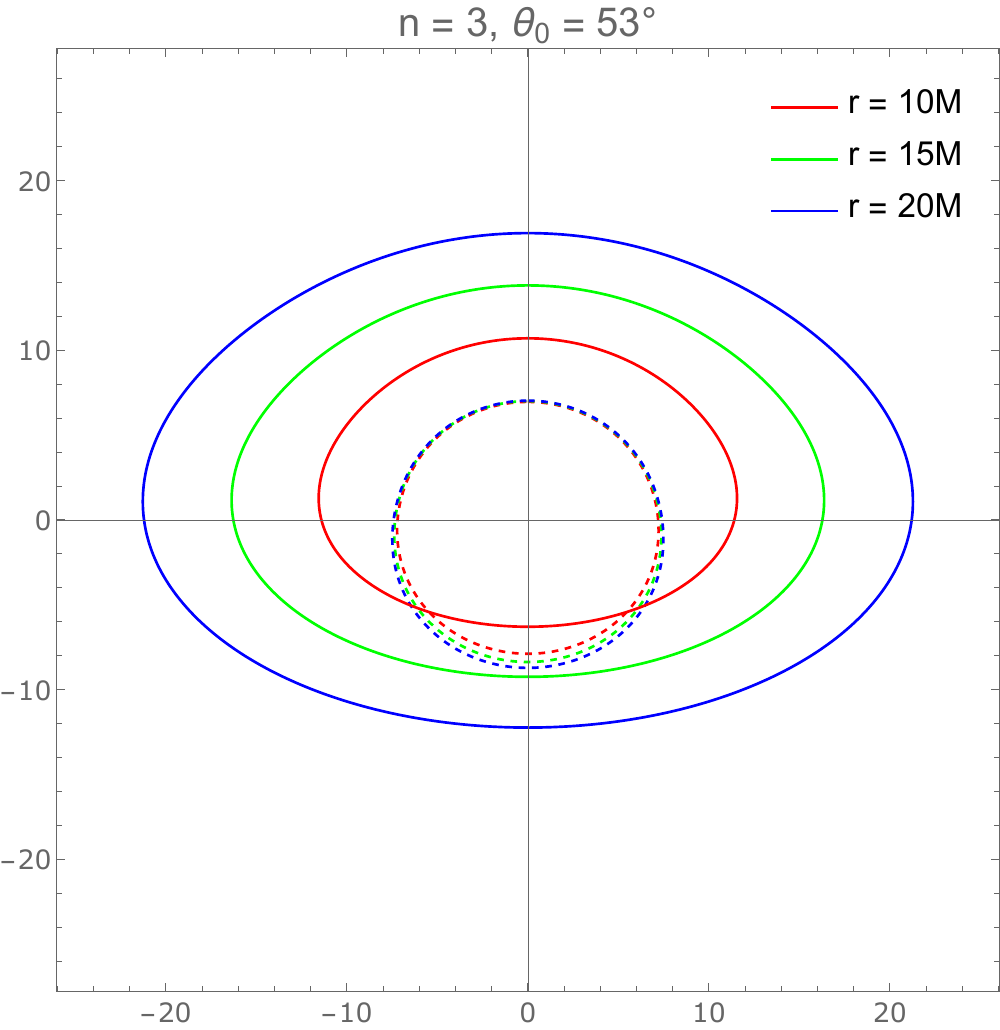}
		\end{minipage}
		\begin{minipage}{0.3\textwidth}
			\centering
			\includegraphics[width=1\textwidth]{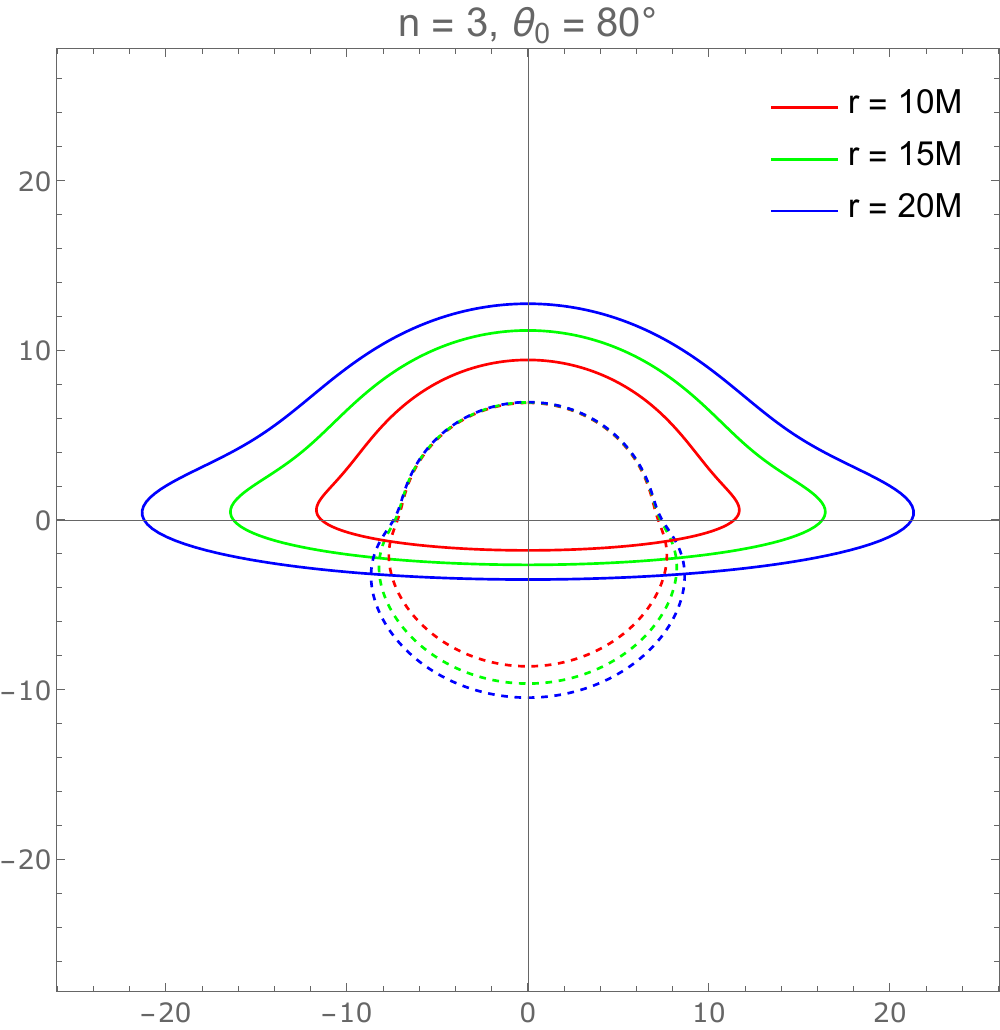}
		\end{minipage}
		\caption{The direct image (solid line) and secondary image (dashed line) when the observer and the accretion disk are located on the same side of the throat. Top panel: Schwarzschild black hole. Middle panel: EB wormhole with $n=2$. Bottem panel: EB wormhole with $n=3$.}
		\label{t6}
	\end{figure*}
	
	\begin{figure*}
		\begin{minipage}{0.3\textwidth}
			\centering
			\includegraphics[width=1\textwidth]{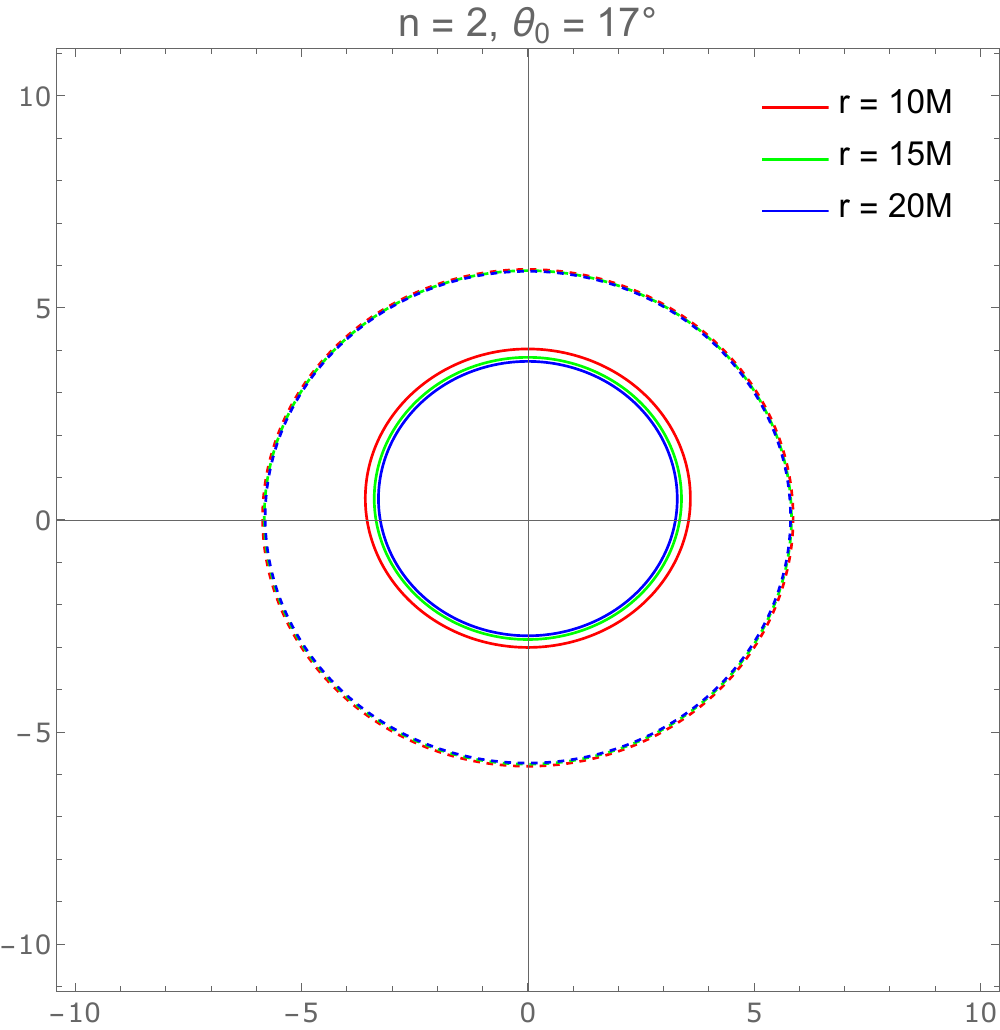}
		\end{minipage}
		\begin{minipage}{0.3\textwidth}
			\centering
			\includegraphics[width=1\textwidth]{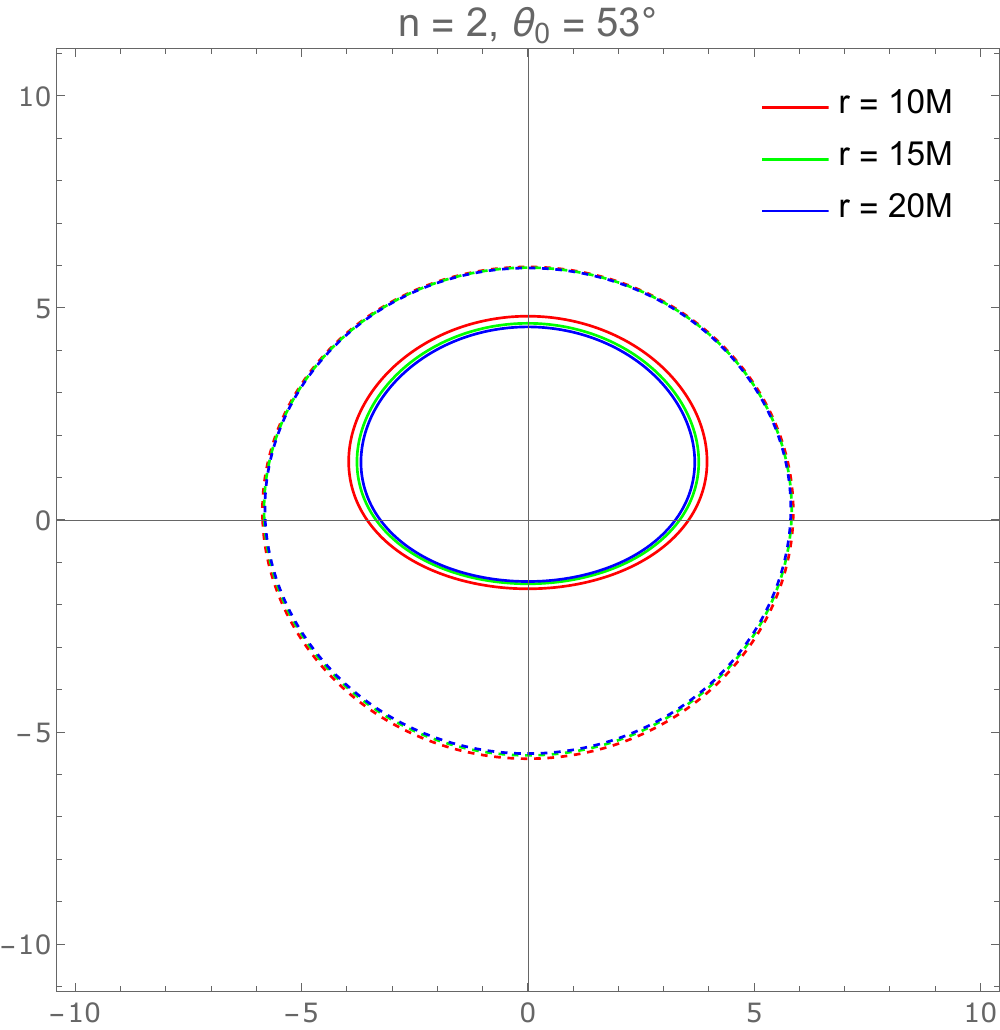}
		\end{minipage}
		\begin{minipage}{0.3\textwidth}
			\centering
			\includegraphics[width=1\textwidth]{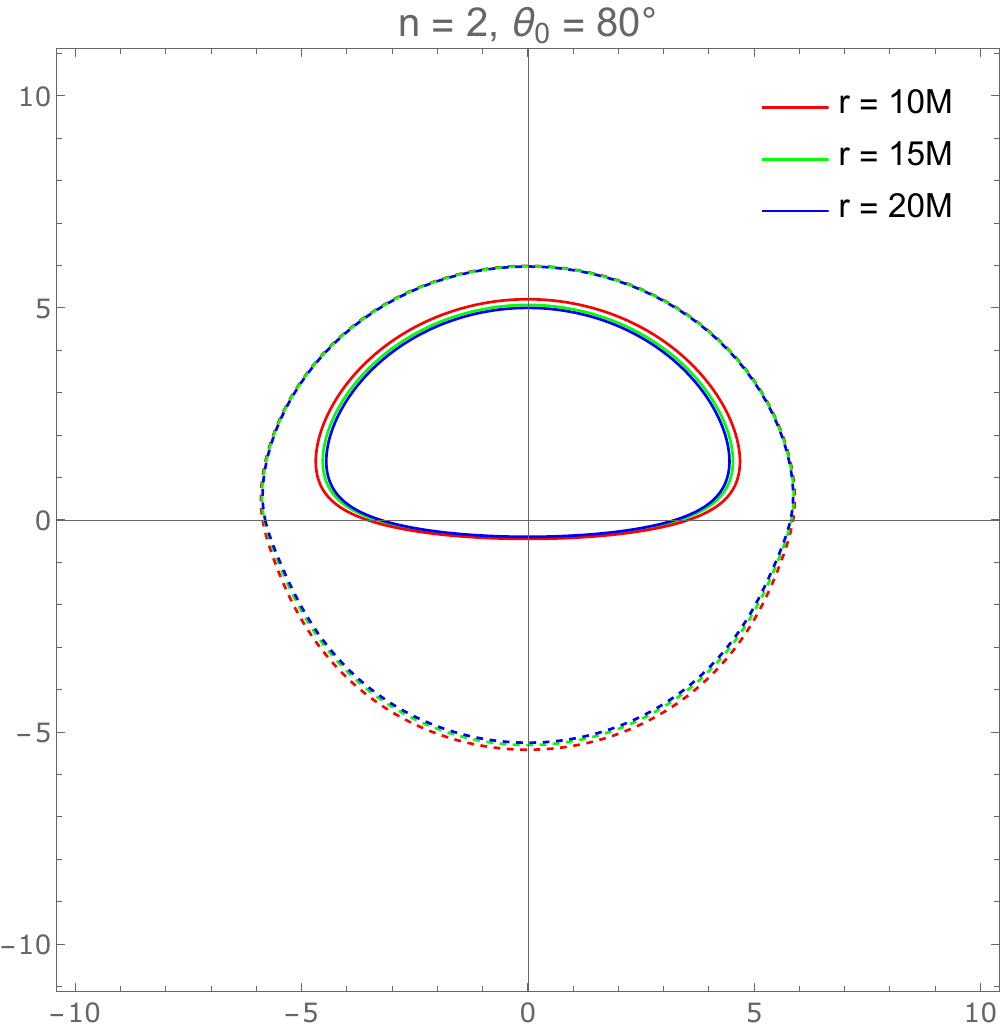}
		\end{minipage}
		\begin{minipage}{0.3\textwidth}
			\centering
			\includegraphics[width=1\textwidth]{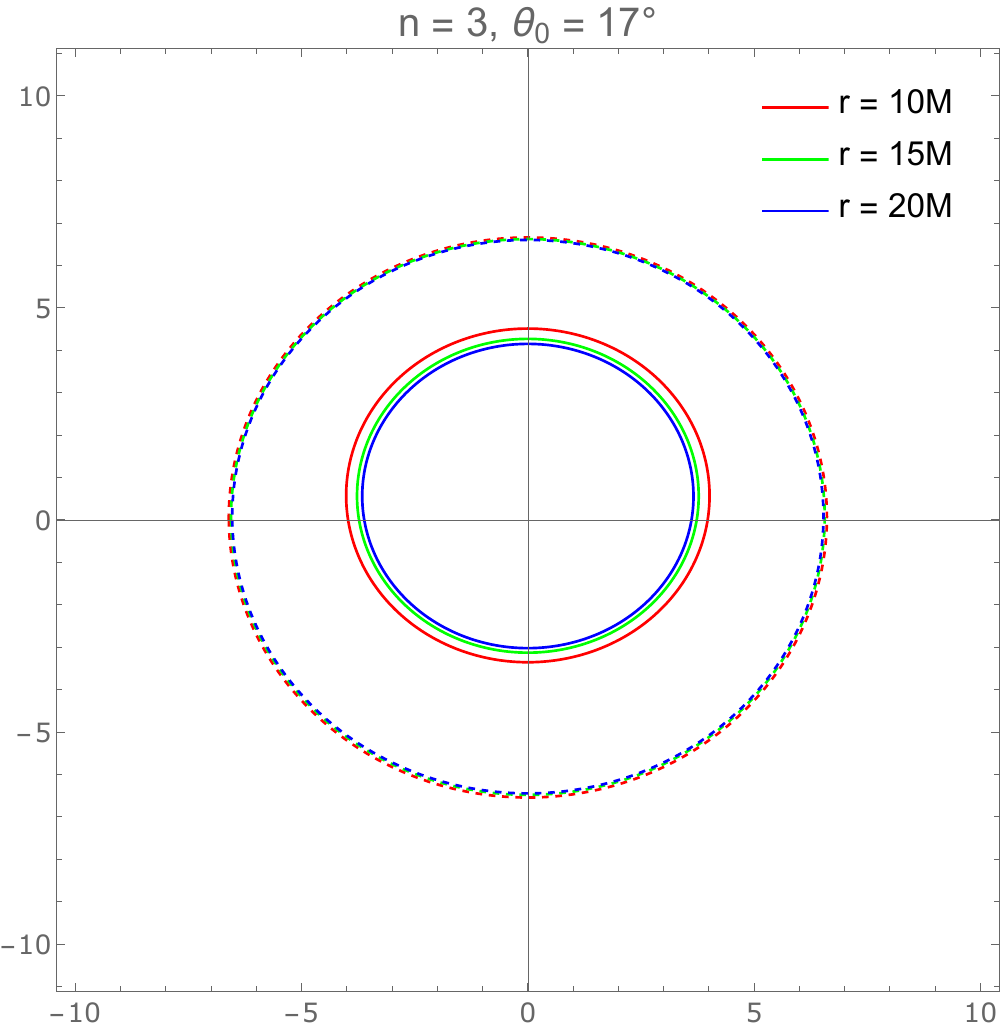}
		\end{minipage}
		\begin{minipage}{0.3\textwidth}
			\centering
			\includegraphics[width=1\textwidth]{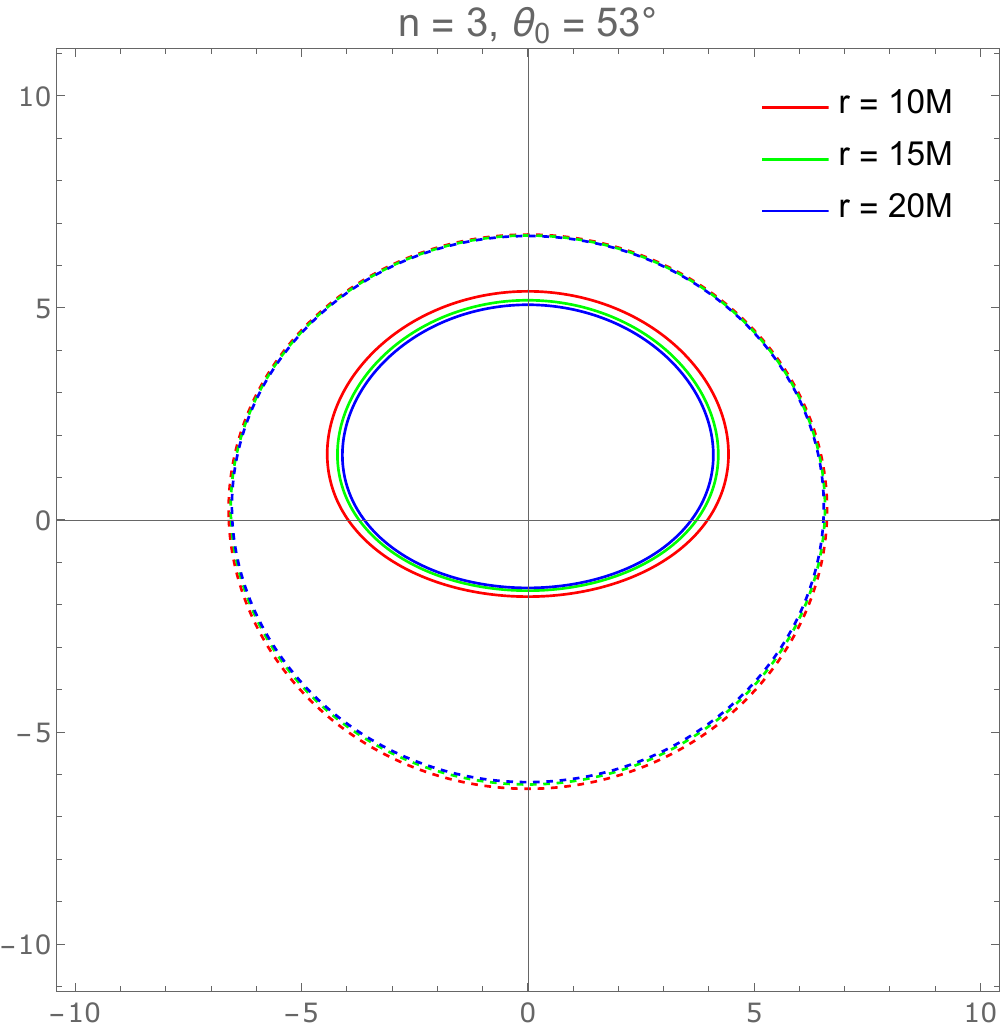}
		\end{minipage}
		\begin{minipage}{0.3\textwidth}
			\centering
			\includegraphics[width=1\textwidth]{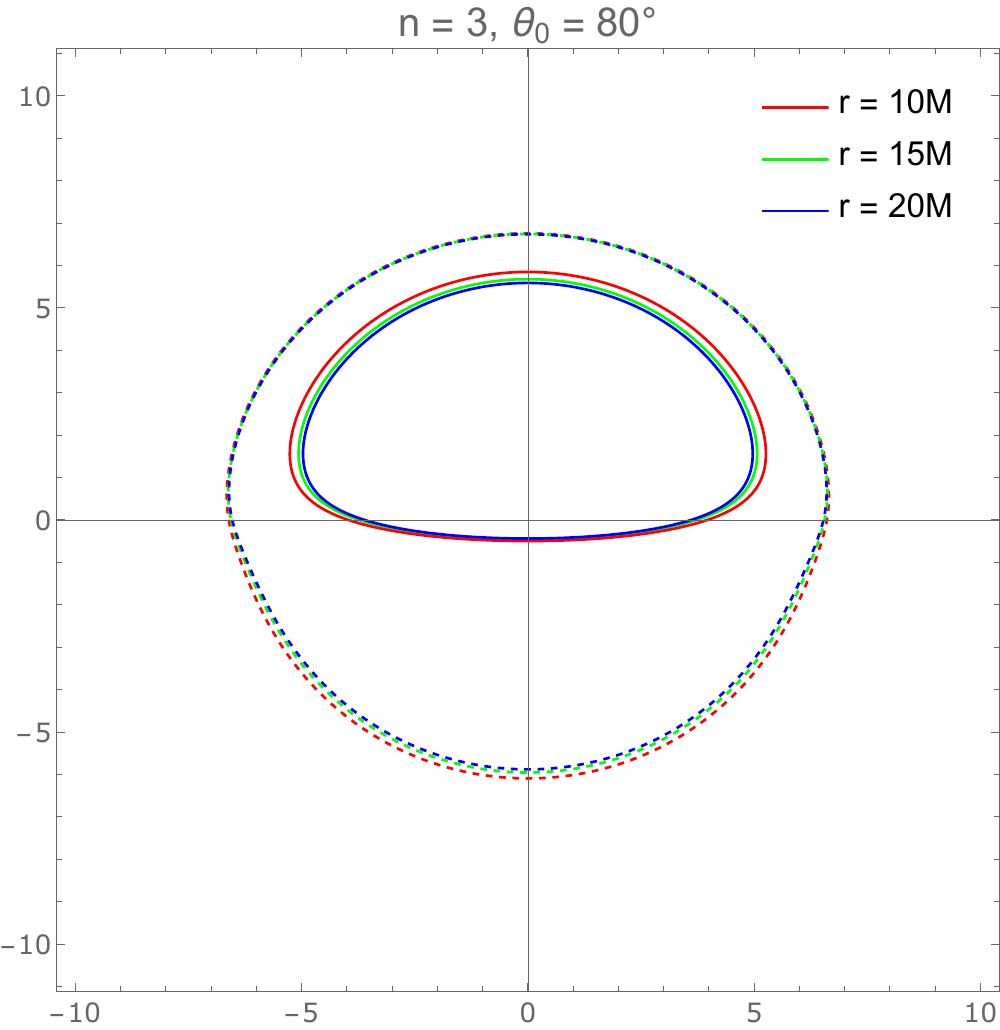}
		\end{minipage}
		\begin{minipage}{0.3\textwidth}
			\centering
			\includegraphics[width=1\textwidth]{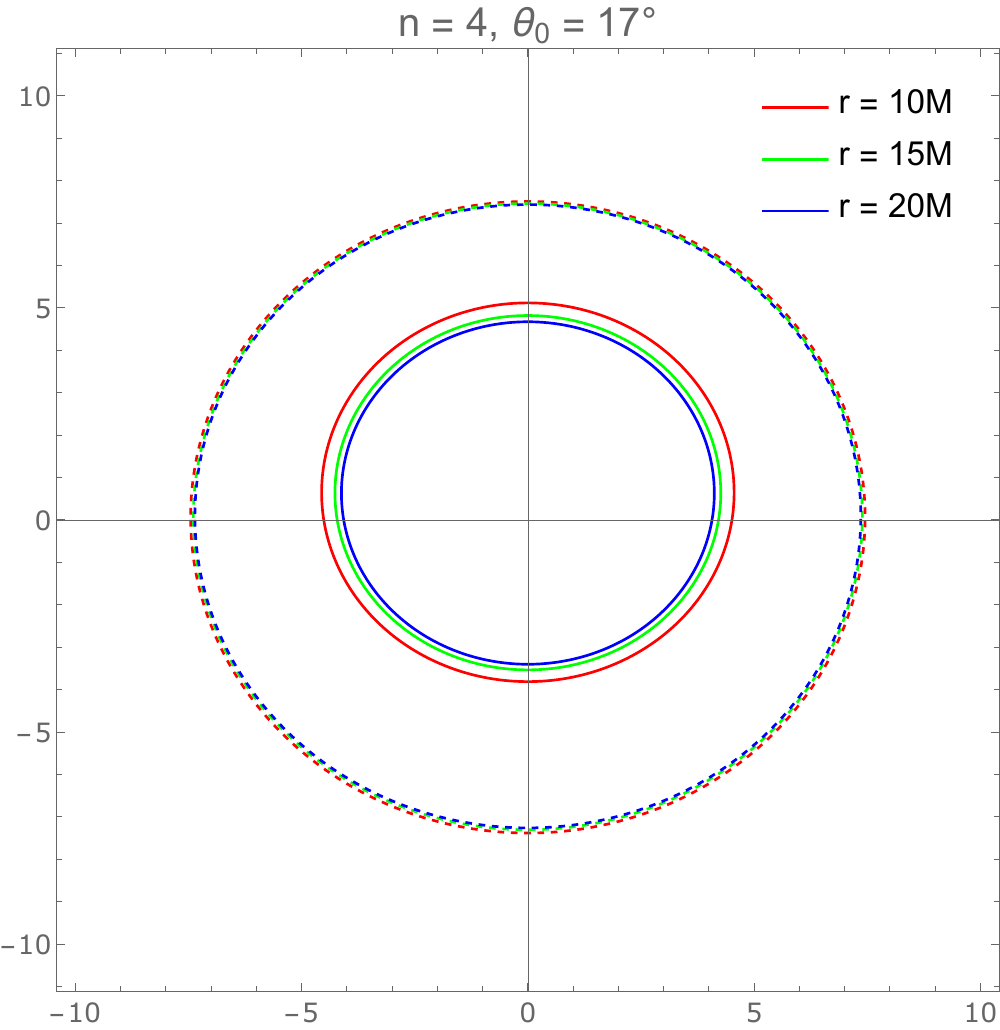}
		\end{minipage}
		\begin{minipage}{0.3\textwidth}
			\centering
			\includegraphics[width=1\textwidth]{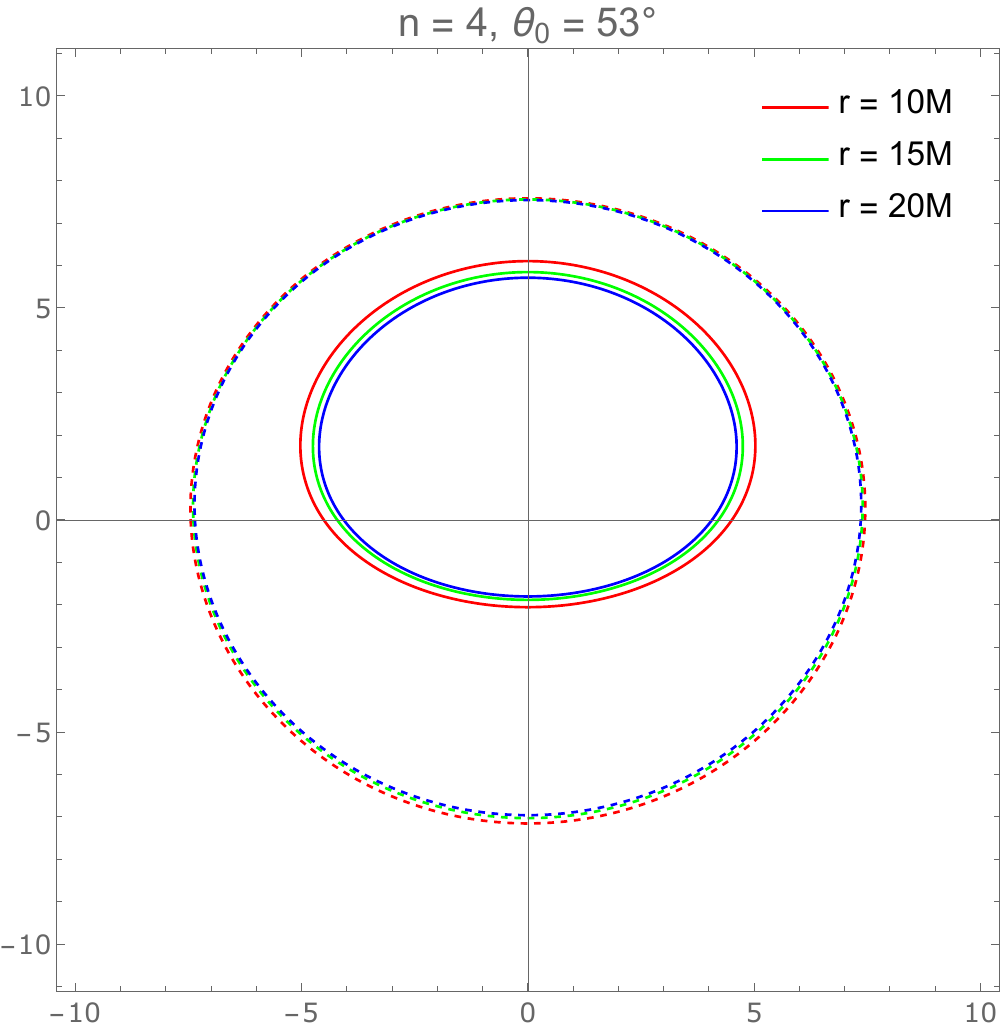}
		\end{minipage}
		\begin{minipage}{0.3\textwidth}
			\centering
			\includegraphics[width=1\textwidth]{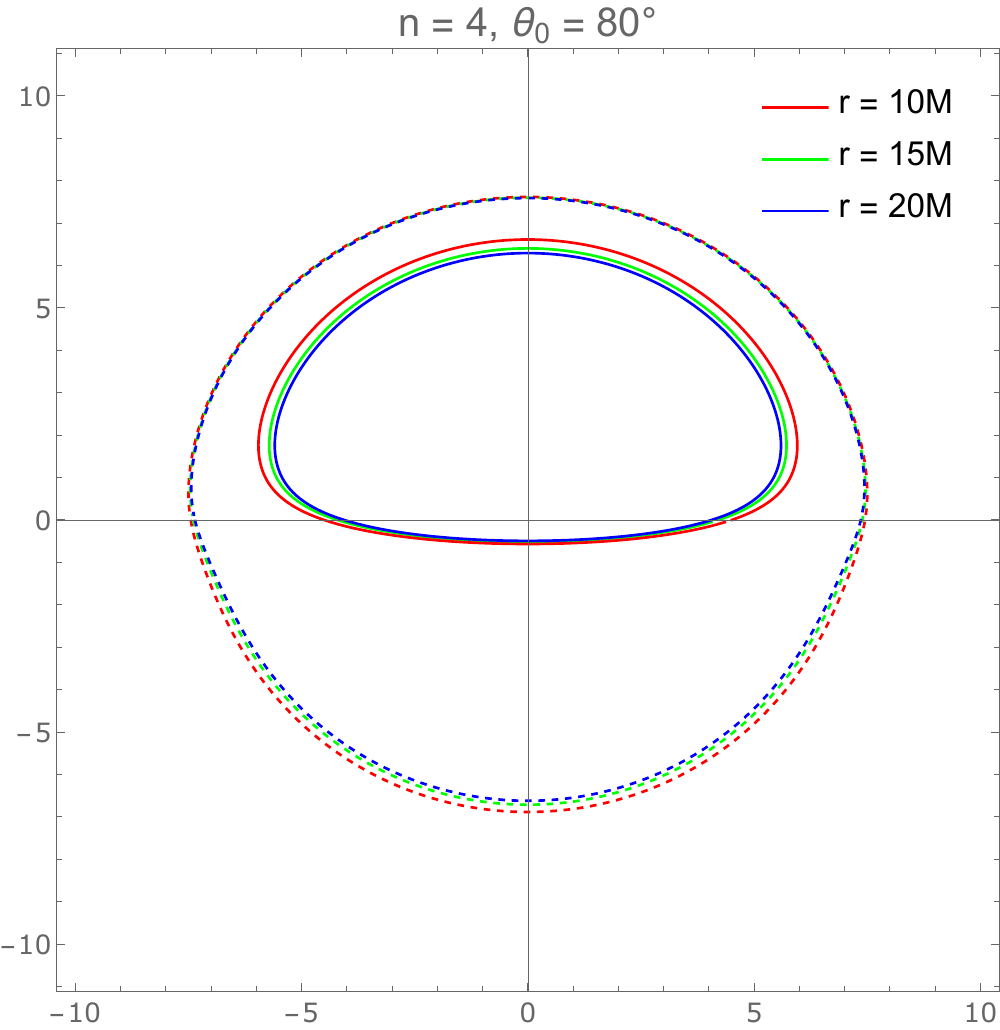}
		\end{minipage}
		\caption{The direct image (solid line) and secondary image (dashed line) when the observer and the accretion disk are located on opposite sides of the throat. Top panel: EB wormhole with $n=2$. Middle panel: EB wormhole with $n=3$. Bottem panel: EB wormhole with $n=4$.}
		\label{t7}
	\end{figure*}
	
	Integrating $r$ in Eq. (12), we can obtain the deflection angle of the photon as
	\begin{equation}\label{25}
		\gamma_1=\int^{+\infty}_{r_{source}}\frac{dr}{\sqrt{\rho(r)\left(\frac{\rho(r)}{b^2}-f(r)\right)}},
	\end{equation}
	in which $r_{source}$ denotes radial coordinate of point $M$. Furthermore, the deflection angle of the photon after passing through the turning point is:
	\begin{equation}\label{26}
		\begin{split}
			\gamma_2 &=2\int^{+\infty}_{r_p}\frac{dr}{\sqrt{\rho(r)\left(\frac{\rho(r)}{b^2}-f(r)\right)}}\\
			&-\int^{+\infty}_{r_{source}}\frac{dr}{\sqrt{\rho(r)\left(\frac{\rho(r)}{b^2}-f(r)\right)}},
		\end{split}
	\end{equation}
	where $r_p$ represents the radial coordinate of the turning point. By combining Eqs. (\ref{24}), (\ref{25}), and (\ref{26}), we have established the relationship between the impact parameter $b$ and $\alpha$. Due to the complexity of the metric (\ref{eq1}), an analytical relation between $\alpha$ and $b$ cannot be derived. Hence, the subsequent calculations are carried out numerically. The ray-tracing calculations were performed using a code based on the backward ray-tracing method described in \cite{23}. Fig. \ref{t6} depicts the direct and secondary images of Schwarzschild black hole and EB wormhole accretion disk as observed in $\mathcal{R}^+$. It can be seen from the figure that the direct images of Schwarzschild black holes and EB wormholes show negligible differences, whereas the secondary images of EB wormholes are slightly smaller than those of Schwarzschild black holes. Furthermore, as $n$ increases, the discrepancy between EB wormholes and Schwarzschild black holes gradually widens. In addition, as $\theta_0$ increases, the deformation of both the direct and secondary images becomes increasingly pronounced.
	
	Fig. \ref{t7} shows the direct image and the secondary image of the EB wormhole when the observer is located in $\mathcal{R}^-$. We can observe that as $n$ increases, both the direct image and the secondary image will significantly enlarge. Additionally, light emitted from positions with smaller radii will produce larger images, due to the position with smaller radius is closer to the observer in the other side of the throat.
	
	\subsection{The radiation flux of the accretion disk}
	We consider the EB wormhole surrounded by a optically thick and geometrically thin accretion disk \cite{57}. In contrast to optically thin accretion disks, optically thick accretion disks have a sufficiently large optical depth that light cannot penetrate the disk. The radiative flux of a standard optically thick accretion disk can be expressed as \cite{57}:
	\begin{equation}
		F=-\frac{\dot{M}}{4\pi \sqrt{-g}}\frac{\Omega^*_{,r}}{(E^*-\Omega^*L^*)^2}\int^r_{r_{in}}(E^*-\Omega^*L^*)L^*_{,r}dr,
	\end{equation}
	where $E_m$, $L_m$, and $\Omega_m$ respectively represent the energy, angular momentum and angular velocity of the particles on the accretion disk. Under the metric of spherical symmetry $ds^2=g_{tt}dt^2+g_{rr}dr^2+g_{\theta \theta}d\theta^2+g_{\varphi \varphi}d\varphi^2$, one can obtain:
	\begin{align}
		E &=-\frac{g_{tt}}{\sqrt{-g_{tt}-g_{\varphi \varphi}\Omega^2}},\\
		L &=\frac{g_{\varphi\varphi}\Omega}{\sqrt{-g_{tt}-g_{\varphi \varphi}\Omega^2}},
	\end{align}
	and
	\begin{equation}
		\Omega = \frac{d\varphi}{dt}=\sqrt{-\frac{g_{tt,r}}{g_{\varphi\varphi,r}}}.
	\end{equation}
	
	In order to obtain the observed radiative flux, we take into account the effects of gravitational redshift and the doppler effect, one can obtain \cite{58}:
	\begin{equation}
		F_{obs}=\frac{F}{(1+z_{r})^4},
	\end{equation}
	where $z_{r}$ is the redshift, which can be expressed as:
	\begin{align}
		1+z_{r}&=\frac{E_{em}}{E_{obs}}=\frac{p^tu_t(1+\Omega \frac{P^\varphi}{P^t})}{\sqrt{-g_{tt}-g_{\varphi\varphi}\Omega^2}}\nonumber \\
		&=\frac{1+b\Omega\sin{\alpha}\sin{\theta_0}}{\sqrt{-g_{tt}-g_{\varphi \varphi}\Omega^2}},
	\end{align}
	in which $E_{em}$ and $E_{obs}$ represent the energy of the photon as emitted from the accretion disk and upon striking the observation plane, respectively.
	
\begin{figure*}
		\begin{minipage}{1\textwidth}
			\centering
			\includegraphics[width=1\textwidth]{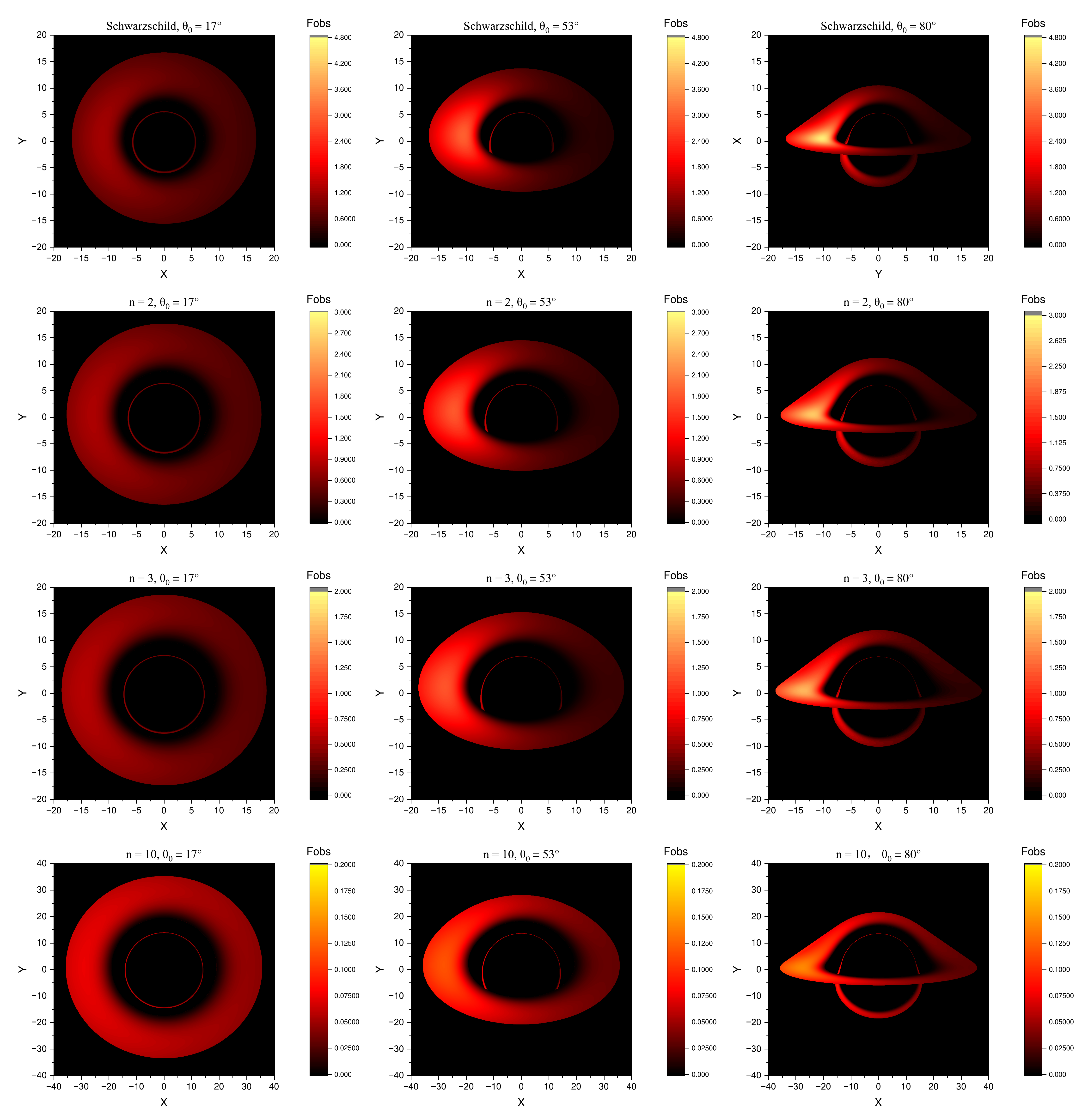}
		\end{minipage}
		\caption{Observational images of Schwarzschild black Holes and EB wormholes under optically thick accretion when the observer and the accretion disk are located on the same side of the throat.}
		\label{t8}
	\end{figure*}

\begin{figure*}
		\begin{minipage}{1\textwidth}
			\centering
			\includegraphics[width=1\textwidth]{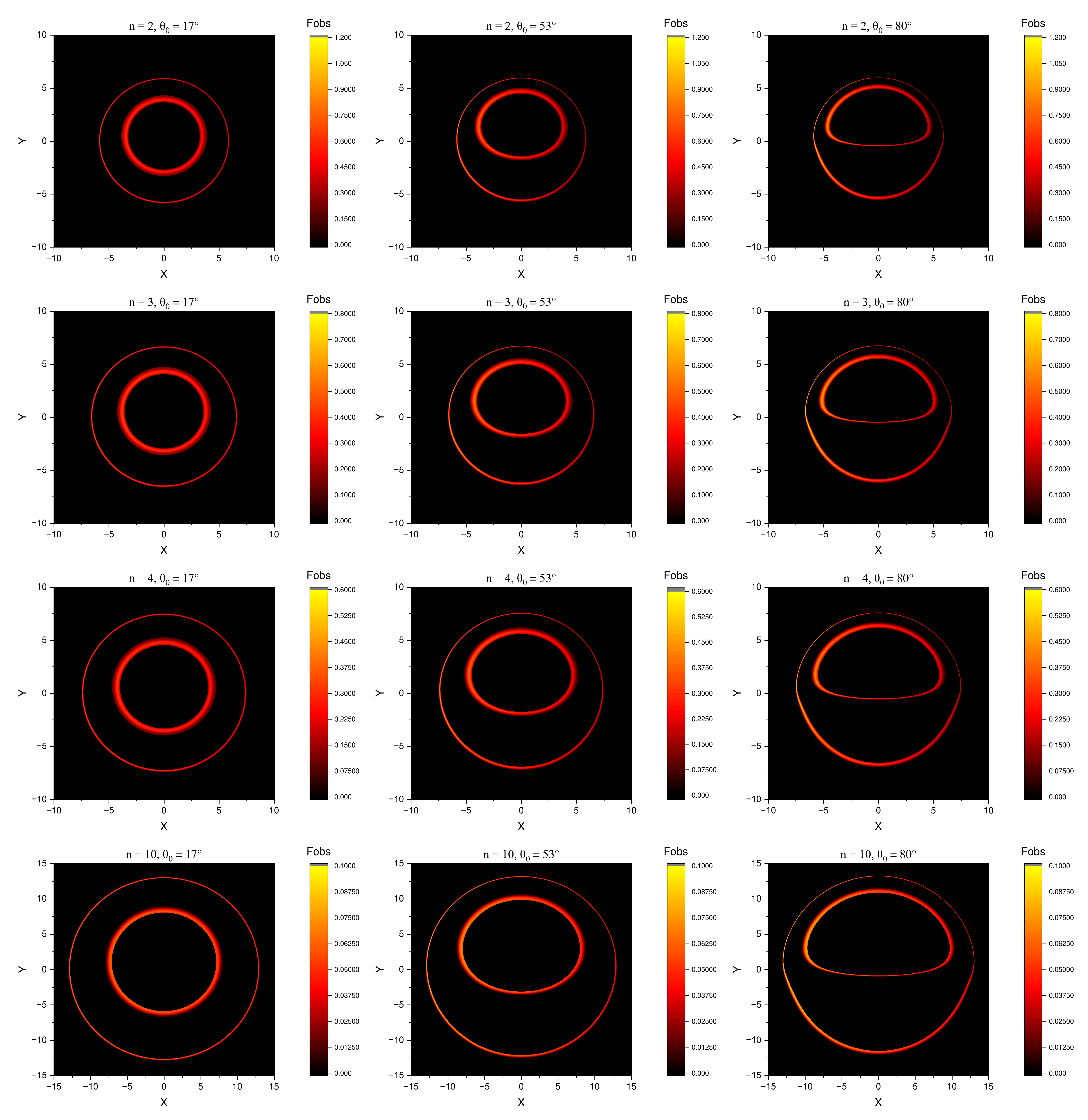}
		\end{minipage}
		\caption{Observational images of Schwarzschild black Holes and EB wormholes under optically thick accretion when the observer and the accretion disk are located on opposite sides of the throat.}
		\label{t9}
	\end{figure*}

	Figs. \ref{t8} and \ref{t9} present simulated images of the optical appearance of Schwarzschild black holes and EB wormholes. In the drawing, we take the maximum intrinsic radiation intensity of the Schwarzschild black hole as the unit, so the radiation intensity in the figure is relative radiation intensity and cannot represent the actual observed flux intensity. It is important to note that due to the relationship described by Eq. (\ref{bwuli}), the images in Fig. \ref{t9} do not represent the actual size of the accretion disk. In reality, the accretion disk in the case of $n=4$ is much larger than that in the case of $n=2$.
	
	From the figure, one can see that when the observer is on the $\mathcal{R}^+$ side (Fig. \ref{t8}), the brightness of the EB wormhole is weaker than that of the Schwarzschild black hole. As $n$ increases, the brightness of the wormhole gradually decreases. However, in terms of morphology, the difference between the wormhole and the black hole is very small. This is mainly because both the wormhole and the black hole act as strong gravitational sources, and their spacetime structures are similar (when only the $\mathcal{R}^+$ side of the wormhole spacetime is considered). As a result, the motion of light rays also exhibits similar characteristics, which is consistent with previous studies comparing different black holes \cite{21,22,23,23j1}. For the situation of observer in $\mathcal{R}^-$ (Fig. \ref{t9}), it can be seen that the shape of the image has undergone a significant change. As $n$ increases, the brightness of the image gradually decreases, while the size of the image gradually increases.
	
\begin{figure*}
    	\begin{minipage}{0.95\textwidth}
    		\centering
    		\includegraphics[width=1\textwidth]{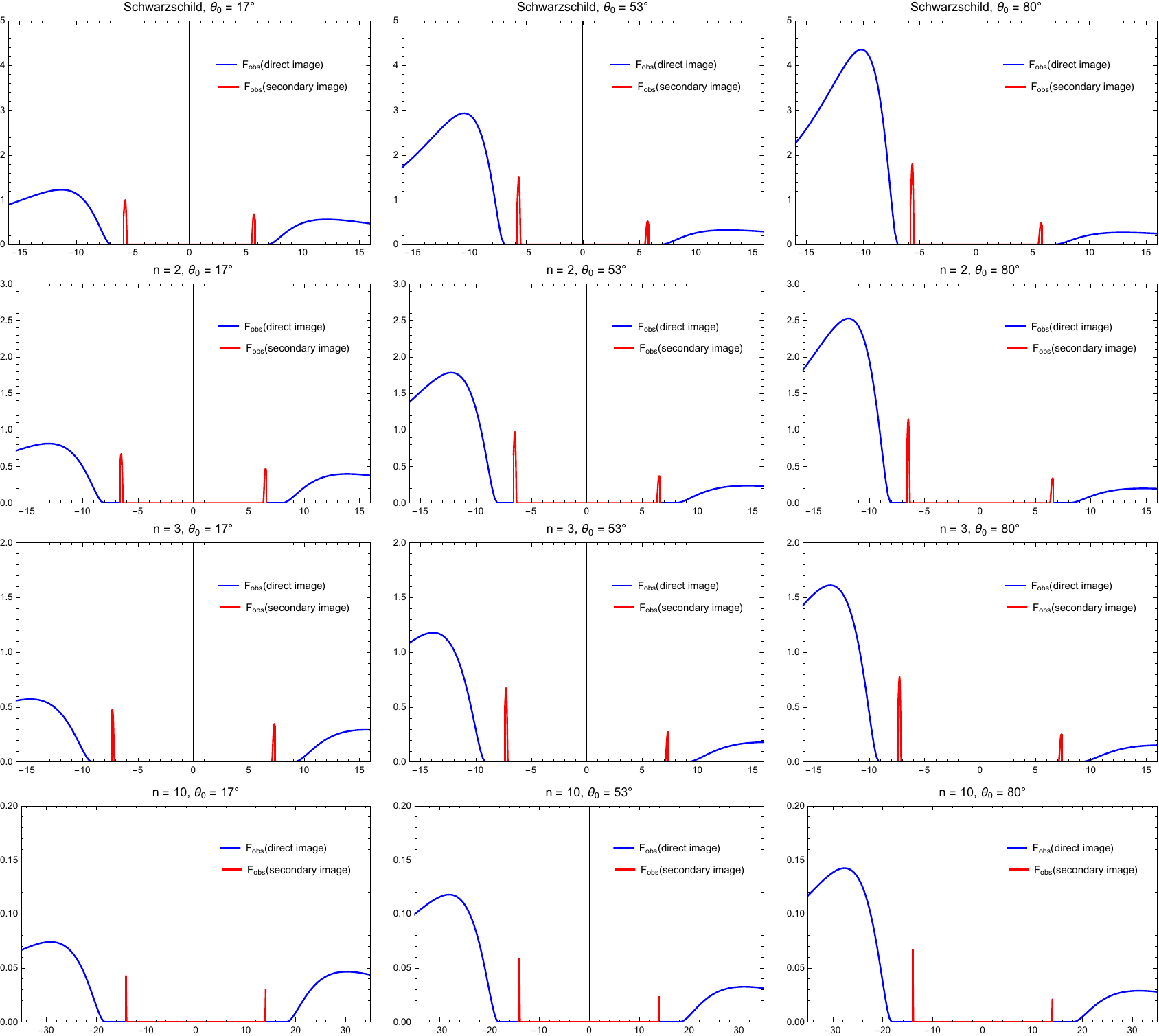}
    	\end{minipage}
    	\caption{Cross-sectional profile of the data in Fig. \ref{t8} along $y=0$. The horizontal axis represents $X$, and the vertical axis represents $F_{obs}$.}
    	\label{tpou1}
    \end{figure*}

\begin{figure*}
		\begin{minipage}{0.95\textwidth}
			\centering
			\includegraphics[width=1\textwidth]{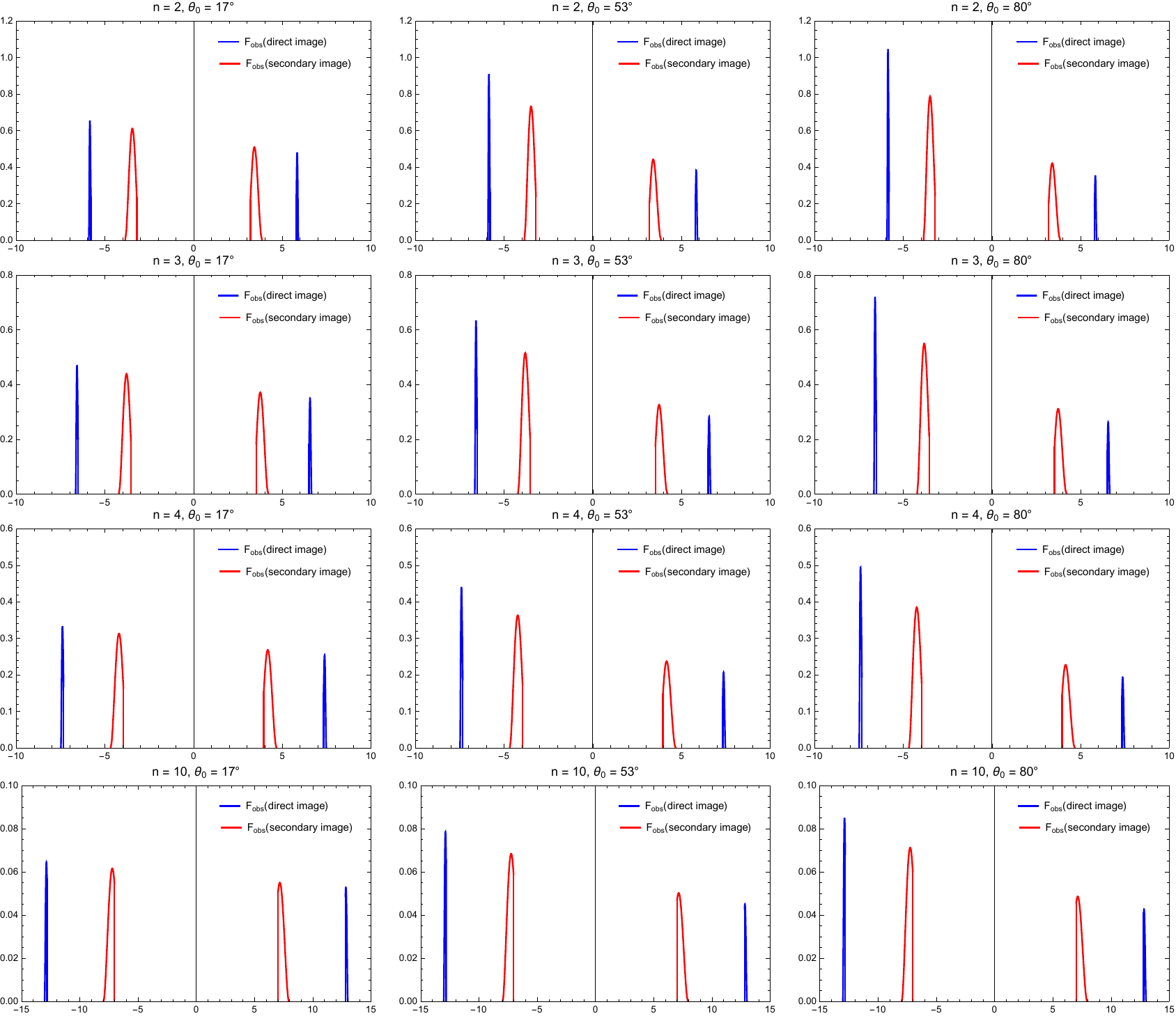}
		\end{minipage}
		\caption{Cross-sectional profile of the data in Fig. \ref{t9} along $y=0$. The horizontal axis represents $X$, and the vertical axis represents $F_{obs}$.}
		\label{tpou2}
	\end{figure*}

	For a clearer comparison of the observational results under different parameters, we present the observed intensity distribution along the slice $y=0$ in Figs. \ref{tpou1} and \ref{tpou2}. It is worth noting that the choice of other $y=constant$ slice does not affect the discussion; here we selected $y=0$, where the secondary image is not obscured. We can be observed from Figs. \ref{tpou1} and \ref{tpou2} that $F_{obs}$ is asymmetric along the X-axis, and the asymmetry gradually increases with increasing $\theta_0$. This is mainly caused by the fact that the larger the observation inclination angle, the smaller the angle between the line of sight and the rotation direction of the accretion disk. Additionally, it can be seen that when $|X|$ is small, the values of $F_{obs}$ for both the direct image and the secondary image are zero. This behavior arises because the corresponding photon trajectories traverse the throat and enter the $\mathcal{R}^-$ region directly, rather than intersecting the accretion disk. Consequently, no radiation is emitted along these trajectories, resulting in a vanishing observed flux.
	 
	By comparing the images of the Schwarzschild black hole and EB wormholes with different $n$ in Figs. \ref{tpou1} and \ref{tpou2}, we can observe that when the observer is located in $\mathcal{R}^+$, the image of the Schwarzschild black hole is very similar to that of the wormhole with a small value of $n$  As $n$ increases, the size of the wormhole image gradually enlarges while its brightness decreases. This behavior is mainly associated with the gradual increase in $r_{isco}$ and the corresponding decrease in the angular velocity of the accretion disk. A similar trend is also observed for the image when the observer is on the other side. The most crucial finding arises in the case where the observer is located in $\mathcal{R}^-$. Since the inner region of the accretion disk is closer to the observer in this case, it appears at the outer edge in the simulated image. we take the direct image as an example, it can be seen from Fig. \ref{tpou2} that the radiative flux at the outermost part of this direct is zero. This contrasts with the scenario of the observer located in $\mathcal{R}^+$, where the radiative flux vanishes at the innermost part of the direct image.

	
	
	\section{\label{sec:level5}The optical appearance of the optically thin accretion disk}
	In this section, we study the optical appearance of EB wormholes surrounded by optically thin accretion disks. We consider an accretion disk emission function with a relatively slow radiative decay and compare the results with those of Huang et al.
	
	The primary distinction between an optically thin and an optically thick accretion disk lies in the ability of light to traverse the disk multiple times. In the simulated image of a wormhole with optically thick accretion conditions presented in the previous section (Fig. \ref{t8}), it can be observed that the secondary images are obscured by the direct image rather than superimposed. This occurs because the light rays that form secondary images cannot penetrate the optically thick accretion disk. To account for scenarios where light passes through the accretion disk multiple times, it is necessary to classify the light rays based on the number of orbits they complete around the wormhole. Here, we define the orbital number $n(b)$ of the light ray as the ratio of light ray's deflection angle to $2\pi$. Note that the deflection angle is computed using Eq. (\ref{25}) for rays that pass through the turning point and Eq. (\ref{26}) for those that do not.
	
	We adopt the model of optically thin emission proposed by Gralla et al. \cite{19}, with the accretion disk in the equatorial plane and the observer at the North Pole. In this configuration, the number of times a light ray crosses the accretion disk can be related to its orbital number $n(b)=\frac{\gamma}{2\pi}$, where $\gamma$ denotes the total deflection angle of the photon. That is, when $n(b)=\frac{1}{4}+\frac{1}{2}k,\ k=0,1,2,\ldots$, the light will cross the accretion disk. Gralla et al. referred to the images formed by the light passing through the accretion disk once, twice, and three times as direct image, lensing ring, and photon ring, respectively. The ranges of the direct image, lensing ring, and photon ring are obtained by solving the equation $n(b)>\frac{1}{4}+\frac{1}{2}k,\ k=0,1,2$ and are summarized in Table 1, where Case 1 indicates that the observer is in $\mathcal{R}^+$, while Case 2 indicates that the observer is in $\mathcal{R}^-$. It can be seen that in Case 2, there is an upper limit to the range of $b$, as the light must traverse the peak of the effective potential to reach the observer, requiring $b<b_c$.
	
	\begin{table}
		\caption{The range of the impact parameter $b$ corresponding to direct image, lensing ring, and photon ring of EB wormhole. $n$ is set to $1.25$.}
		\label{tt11}
		\begin{ruledtabular}
			\begin{tabular}{cccccc}
				\\[-7pt]
				& Direct image & Lensing ring & Photon ring \\
				\\[-7pt]
				\hline
				\\[-6pt]
				Case 1 & $(3.358,+\infty)$ & $(5.425,6.516)$ & $(5.554,5.591)$ \\[1pt]
				Case 2 & $(3.358,b_c)$ & $(5.425,b_c)$ & $(5.554,b_c)$ \\[1pt]
			\end{tabular}
		\end{ruledtabular}
	\end{table}
	
	\begin{figure*}
		\begin{minipage}{0.8\textwidth}
			\centering
			\includegraphics[width=1\textwidth]{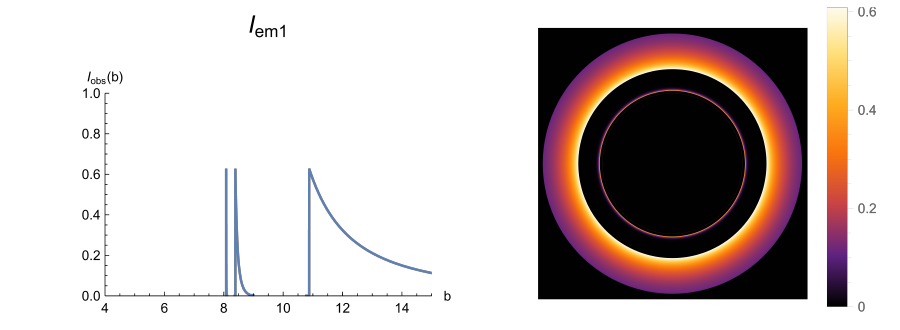}
		\end{minipage}
		\begin{minipage}{0.8\textwidth}
			\centering
			\includegraphics[width=1\textwidth]{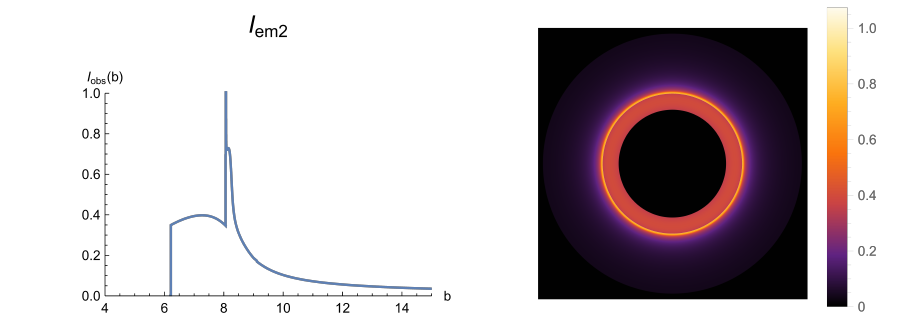}
		\end{minipage}
		\caption{Observational radiation flux (left) and optical images (right) of EB wormholes in optically thin accretion from the $\mathcal{R}^+$ side. The observer and the accretion disk are on the same side of the throat. $n$ is set to $4.441$ (consistent with Ref. \cite{53}). Top panel: Emission function $I_{em1}$. Bottom panel: Emission function $I_{em2}$}
		\label{tbaotong}
	\end{figure*}
	
	\begin{figure*}
		\begin{minipage}{0.8\textwidth}
			\centering
			\includegraphics[width=1\textwidth]{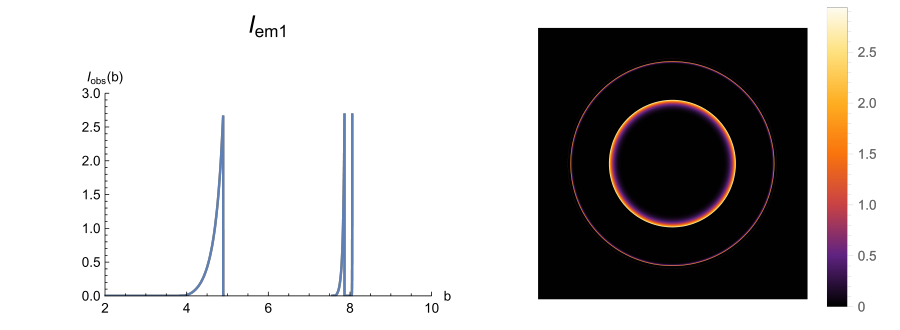}
		\end{minipage}
		\begin{minipage}{0.8\textwidth}
			\centering
			\includegraphics[width=1\textwidth]{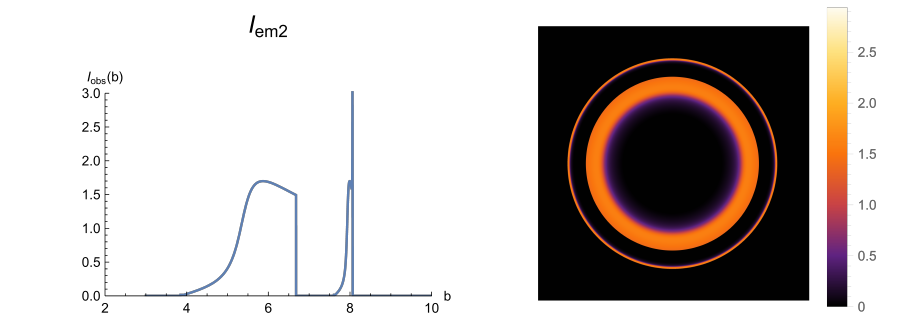}
		\end{minipage}
		\caption{Observational radiation flux (left) and optical images (right) of EB wormholes in optically thin accretion from the $\mathcal{R}^-$ side. The observer and the accretion disk are on the opposite sides of the throat. $n$ is set to $4.441$ (consistent with Ref. \cite{53}). Top panel: Emission function $I_{em1}$. Bottom panel: Emission function $I_{em2}$.}
		\label{tbaoyi}
	\end{figure*}
	
	We suppose the emission function of the two types of accretion disks:
	\begin{equation}
		I_{em1}=\left\{
		\begin{aligned}
			&\frac{(r_{isco}-5)^2}{(r-5)^2},~ &r\geq r_{isco} \\
			&0, ~&r<r_{isco}
		\end{aligned}
		\right.
	\end{equation}
	\begin{equation}
		I_{em2}=\left\{
		\begin{aligned}
			&\frac{\frac{\pi}{2}-\arctan(r-5)}{\frac{\pi}{2}-\arctan(-3)},~ &r\geq r_c \\
			&0, ~&r<r_c
		\end{aligned}
		\right.
	\end{equation}
	in which $I_{em1}$ is obtained by normalizing the radiation function in  \cite{53}, with its inner radiation boundary at $r_{isco}$. The radiation boundary of $I_{em2}$ is located at the photon sphere radius $r_c$, and its radiation decays more slowly as the radius increases. It can be found that this model is a toy model, but it still has value in describing the positions of the photon ring, the lensing ring, and the direct image. Taking the redshift effect into account, the observed radiation flux finally obtained is	
	\begin{equation}
		I_{obs}=\sum f(r)^2 I_{em}|_{r=r_n(b)},
	\end{equation}
	where $r_n(b)$ represents the radial coordinate when light passes through the accretion disk for the $n$~th time, which can be obtained by solving the geodesic equation. 
	
	The observed radiation flux and optical images for observers in $\mathcal{R}^+$ and $\mathcal{R}^-$ are shown in Fig. \ref{tbaotong} and Fig. \ref{tbaoyi}, respectively. From Fig. \ref{tbaotong}, it can be observed that for emission function $I_{em1}$, the direct image, lensing ring, and photon ring each possess distinct boundaries, whereas the case for emission function $I_{em2}$ is different, which primarily arises from the emission boundary being positioned closer to the throat. In the $\mathcal{R}^-$ case (Fig. \ref{tbaoyi}), the radial order of the direct image, lensing ring, and photon ring for emission function $I_{em1}$ is inverted from the inside out, resulting in a notably brighter inner region. A similar phenomenon is observed for emission function $I_{em2}$, where the observed radiation flux increases from zero as the radius increases, representing a radial trend exactly opposite to that shown in Fig. \ref{tbaotong}.

	\section{\label{sec:level6}SUMMARY}
	In this work, we studied the optical appearance of the EB wormhole observed from both sides of the throat, including both optically thick and thin accretion backgrounds. First of all, we analyzed the global structure of the EB wormhole spacetime and discussed the geodesics of photons and particles, calculating important parameters such as the critical impact parameter $b_c$, the radius of the circular orbit of a photon $r_s$, and the innermost stable circular orbit $r_{isco}$. In the analysis of null geodesics, we discovered a connection between the impact parameter $b$ and the aiming distance $d^*$, and provided a proof of this in the Appendix. \ref{app}. Furthermore, the analysis of the timelike geodesics indicates that the accretion disk can only exist in $\mathcal{R}^+$.
	
	In Sec. \ref{sec:level4}, the optical appearance of the EB wormhole with optically thick accretion was studied. We traced the light emitted from the accretion disk of the wormhole and obtained the direct image (Fig. \ref{t6}) and the secondary image (Fig. \ref{t7}). Then we considered gravitational redshift and the Doppler effect, calculated the observed radiation flux, and obtained the simulation image of the optical appearance (Figs. \ref{t8} and \ref{t9}). The results show that when the observer is located in \(\mathcal{R}^+\), the brightness of the wormhole is dimmer than that of the Schwarzschild black hole, and as the parameter \(n\) increases, the brightness gradually decreases while the apparent size of the image gradually increases. In contrast, for the observer in \(\mathcal{R}^-\), the image morphology changes significantly, and increasing \(n\) also leads to a decrease in brightness and an increase in the apparent size of the wormhole image.
	
	In Sec. \ref{sec:level5}, we investigated the optical appearance of the EB wormhole with optically thin accretion under two different emission functions. For emission function $I_{em1}$, our results are consistent with \cite{53}, where the image observed at $\mathcal{R}^-$ (Fig. \ref{tbaoyi}) appears as an inside-out inversion of the image seen at $\mathcal{R}^+$ (Fig. \ref{tbaotong}). A similar behavior occurs for emission function $I_{em2}$: when the observer is in $\mathcal{R}^+$, the observed radiation flux gradually increases from zero from the outside in; conversely, for an observer in $\mathcal{R}^-$, the flux increases from zero from the inside out. Unlike the optically thick case, the direct image of the optically thin accretion disk does not block the lensing ring and the photon ring, thus allowing radiation from regions closer to the horizon to reach the observer plane. At present, the optical thickness of the accretion flow in the M87* black hole image is still under debate. Lu et al. found that the flow appears optically thick at 86 GHz but becomes optically thin at 230 GHz \cite{59}. With its broader frequency coverage and improved angular resolution, the next-generation Event Horizon Telescope is expected to provide stronger observational constraints on the optical-depth structure of the accretion flow, thereby helping to clarify the physical origin of the observed black hole image \cite{60}.
	
	Our results show that if the observer is on the $\mathcal{R}^+$ side, the wormhole with a small parameter $n$ produces an image that is very similar to that of a black hole. However, for large $n$, the $r_{isco}$ of the wormhole gradually increases, resulting in a shadow area in the image that is significantly larger than that of a black hole. On the other hand, if the observer is on the $\mathcal{R}^-$ side, the image differs markedly from that of a black hole. Therefore, for the images taken by the EHT, wormholes with large $n$ and those where the observer is on the $\mathcal{R}^-$ side can be ruled out. A key finding of our study is that, under both optically thick and thin accretion, when the observer is located on the opposite side of the accretion disk, the resulting images exhibit a counterintuitive brightness profile. Since the inner region of the disk is closer to the observer, the interior of a certain rank of image (for instance, the direct image in the case of optical thick accretion) corresponds to the outer side of the accretion disk, while the exterior of the image corresponds to the inner side of the accretion disk. That is, the image observed on the $\mathcal{R}^-$ side is similar to the internal and external inversion of the image on the $\mathcal{R}^+$ side, which is particularly evident in the case of thin optics, as can be seen, the radiation intensity of direct image radiation in Fig. \ref{tbaoyi} gradually intensifies from the inner part to the outer part, which is exactly the opposite of Fig. \ref{tbaotong}. This distinctive brightness signature could serve as a potential observational fingerprint for identifying wormholes and may guide future electromagnetic observations of black hole candidates.

	\begin{acknowledgments}
		This work was supported by the National Key R\&D Program of China (Grant No. 2023YFA1607902), the National Natural Science Foundation of China (Grant Nos. 12173031, 12221003, 12505060, and 12494572), the Fund of National Key Laboratory of Plasma Physics (Grant No. 6142A04240201), the China Manned Space Program (Grant No. CMS-CSST-2025-A13), the Fund Project of Chongqing Normal University (Grant No: 24XLB033), the Chongqing Natural Science Foundation General Program (Grant No. CSTB2025NSCQ-GPX1019), and Fapesq-PB of Brazil.
	\end{acknowledgments}
	
	\appendix
	
	\section{THE PHYSICAL MEANING OF IMPACT PARAMETER $b$}\label{app}
	Here, we discuss in detail the physical meaning of the impact parameter $b$ in the general static spherically symmetric spacetime:
	\begin{equation}
		ds^2=-h(r)dt^2+\frac{1}{f(r)}dr^2+\rho(r)(d\theta^2+\sin^2{\theta}d\varphi^2).
	\end{equation}
	
	Expanding the definition of $b$, we can obtain:
	\begin{equation}\label{A2}
		b=\frac{L}{E}=\frac{\rho(r)\dot{\varphi}}{-h(r)\dot{t}}=\frac{\rho(r)d\varphi}{-h(r)dt}=\frac{\rho(r)}{-h(r)}\frac{d\varphi}{dr}\frac{dr}{dt}.
	\end{equation}
	
	When $r\rightarrow \pm\infty$, we have $\varphi=\arctan{d^*/r}\approx d^*/r$, in which $d^*$ is the aiming distance of the light ray at infinity. Thus, Eq. (\ref{A2}) can be expressed as
	\begin{equation}\label{A3}
		b=\frac{\rho(r)}{-h(r)}\frac{d\varphi}{dr}\frac{dr}{dt}\approx \frac{\rho(r)}{-h(r)}\frac{d\left(\frac{d^*}{r}\right)}{dr}\frac{dr}{dt}=\frac{\rho(r)}{h(r)}\frac{d^*}{r^2}\frac{dr}{dt}.
	\end{equation}
	
	According to the null geodesic equation $ds^2=0$, one can obtain:
	\begin{equation}\label{A4}
		-1+\frac{\left(\frac{dr}{dt}\right)^2}{f(r)h(r)}+\frac{\rho(r)}{h(r)}\left(\frac{d\varphi}{dt}\right)^2=0.
	\end{equation}
	
	By combining Eqs. (\ref{A2}) and (\ref{A4}), we can get
	\begin{equation}\label{A5}
		-1+\frac{\left(\frac{dr}{dt}\right)^2}{f(r)h(r)}+\frac{h(r)b^2}{\rho(r)}=0,
	\end{equation}
	where the last term on the left side of the equation approaches zero as $r\rightarrow \pm \infty$. Therefore, one can obtain:
	\begin{equation}\label{A6}
		\frac{dr}{dt}=\sqrt{f(r)h(r)}.
	\end{equation}
	
	Note that Eq. (\ref{A6}) only holds when $r\rightarrow \pm \infty$. By combining Eq. (\ref{A3}) and Eq. (\ref{A6}), We can obtain the relationship between the impact parameter $b$ and the aiming distance $d$:
	\begin{equation}\label{A7}
		b=\lim_{r\rightarrow \pm \infty}\frac{\rho(r)}{r^2}\sqrt{\frac{f(r)}{h(r)}}d^*.
	\end{equation}
	
	For the EB wormhole described by Eq. (\ref{eq1}), there is
	\begin{equation}\label{A8}
		b=\lim_{r\rightarrow \pm \infty}\frac{d^*}{f(r)}.
	\end{equation}

\end{document}